\documentclass[]{article}
\usepackage[a4paper, left=2.4cm, right=2.4cm, top=2.4cm, bottom=2.4cm]{geometry}

\usepackage{xcolor}
\usepackage{amsmath,amssymb}
\usepackage{svg}
\usepackage{overpic}
\usepackage{tabularx}
\usepackage{float}
\usepackage{listliketab}
\usepackage{lscape}
\usepackage{caption}
\usepackage{standalone}
\usepackage{tikz}
\usepackage{xurl}
\usetikzlibrary {arrows.meta}
\usetikzlibrary{shapes.geometric}
\usepackage{graphicx}
\usepackage{setspace}
\usepackage[T1]{fontenc}
\usepackage[utf8]{inputenc}
\usepackage{textcomp} 
\usepackage{adjustbox}

\usepackage[scaled]{helvet}

\usepackage{longtable,booktabs,array}
 
\usepackage{calc}
\usepackage{etoolbox}

\usepackage[style=apa,backend=biber]{biblatex}
\DeclareLanguageMapping{english}{english-apa}
\let\cite\parencite

\def\fps@figure{htbp} % Set default figure placement to htbp

\usepackage{bookmark}
\author{}
\date{}

\begin{document}

\begin{Large}
{A Systematic Review of Spatio-Temporal Statistical Models: Theory, Structure, and Applications}
\end{Large}

\textbf{Article Category: Systematic Review}

\textbf{Authors:}

{\def\LTcaptype{none} % do not increment counter
\begin{longtable}[]{@{}
  >{\raggedright\arraybackslash}p{(\linewidth - 0\tabcolsep) * \real{0.8994}}@{}}
\toprule\noalign{}
\begin{minipage}[b]{\linewidth}\raggedright
\textbf{Isabella Habereder}

Chair of Statistics, Georg-August University Göttingen, i.habereder@media-bias-research.org, \url{https://orcid.org/0009-0001-8708-2314}

\end{minipage} \\
\midrule\noalign{}
\endhead
\bottomrule\noalign{}
\endlastfoot
\textbf{Thomas Kneib}

Chair of Statistics, Georg-August University Göttingen, tkneib@uni-goettingen.de; \url{https://orcid.org/0000-0003-3390-0972} \\
\midrule\noalign{}

\textbf{Isao Echizen}

Information and Society Research Division, National Institute of Informatics (NII), iechizen@nii.ac.jp; \url{https://orcid.org/0000-0003-4908-1860} \\
\midrule\noalign{}

\textbf{Timo Spinde}

Information and Society Research Division, National Institute of Informatics (NII), t.spinde@media-bias-research.org; \url{https://orcid.org/0000-0003-3471-4127} \\
\end{longtable}
}

\textbf{Abstract}

Data with spatio-temporal attributes are prevalent across many research fields, and statistical models for analyzing spatio-temporal relationships are widely used. Existing reviews focus either on specific domains or model types, creating a gap in comprehensive, cross-disciplinary overviews. To address this, we conducted a systematic literature review following the PRISMA guidelines, searched two databases for the years 2021-2025, and identified 83 publications that met our criteria. We propose a classification scheme for spatio-temporal model structures and highlight their application in the most common fields: epidemiology, ecology, public health, economics, and criminology. Although tasks vary by domain, many models share similarities. We found that hierarchical models are the most frequently used, and most models incorporate additive components to account for spatio-temporal dependencies. The preferred model structures differ among fields of application. We also observe that research efforts are concentrated in only a few specific disciplines, despite the broader relevance of spatio-temporal data. Furthermore, we notice that reproducibility remains limited. Our review, therefore, not only offers inspiration for comparing model structures in an interdisciplinary manner but also highlights opportunities for greater transparency, accessibility, and cross-domain knowledge transfer.

\section{Introduction}\label{sec:introduction}

Spatio-temporal data\footnote{By spatio-temporal data, we refer to data with attributes of both time and space \cite{cressie2011statistics}} arise in a wide range of application areas, including epidemiology \cite{amaral_spatio-temporal_2023}, environmental research \cite{choi_short-term_2021}, economics \cite{elhorst_spatial_2022}, social sciences \cite{Spinde2025}, and health \cite{beloconi_spatio-temporal_2021}. Such data are generated by sensors, digital media platforms, socioeconomic surveys, and environmental monitoring systems, all of which record measurements across space and time \cite{cressie2011statistics}.

The diversity of fields in which spatio-temporal data occur gives rise to a broad spectrum of tasks for advanced statistical modeling. Typical applications include modeling the spread of infectious diseases \cite{amaral_spatio-temporal_2023, ngwira_spatial_2021, tam_bayesian_2024} and assessing the health effects of air pollution \cite{beloconi_spatio-temporal_2021, martenies_spatiotemporal_2021, saez_spatial_2022}. We refer to the statistical models used for these purposes as spatio-temporal models. A spatio-temporal statistical model is a mathematical framework that describes and analyzes data that vary across space and time \cite{cressie2011statistics}.

The challenge lies in specifying the structure of these models to ensure that relationships across spatial units and time points adequately reflect the underlying phenomenon. Several categories of spatio-temporal models exist, including regression models, point processes, and stochastic compartment models, each encompassing distinct modeling strategies. Researchers make a common distinction between descriptive (in this context, i.e., covariance-based) approaches and dynamic processes based on conditional probability \cite{stroud2001dynamic, wikle2015modern, wikle2019spatio}. 

Despite the widespread use of these methods, comprehensive reviews remain scarce. One foundational work in the field is \cite{cressie2011statistics}, which provides a theoretically rigorous classification of spatio-temporal processes based on data structure and methodology. However, this treatment is highly technical and includes only limited practical examples. Other reviews focus on a single domain (e.g., \cite{aswi_bayesian_2019, byun_systematic_2021, martinez-minaya_species_2018, odhiambo_spatial_2023}) or on a specific model class, such as point processes (e.g., \cite{gonzalez_spatio-temporal_2016, reinhart_review_2018}). Furthermore, we note a discrepancy between the models described in textbooks and theoretical overview works and the actual model structures used in practice.
Consequently, there is a lack of up-to-date systematic overviews (see Section \ref{sec:related_work}) that integrate perspectives across model types and application areas while also providing a basic classification scheme for the used spatio-temporal model structures. 

This work aims to present a structured and comparative overview of spatio-temporal model structures and to discuss their applications across disciplines. Such a literature review is overdue, and given the widespread use of spatio-temporal methods, it can benefit various domains by increasing awareness of available techniques and challenges, as well as summarizing the latest state-of-the-art developments. We address the following research questions:

(RQ1) What are the most commonly used spatio-temporal model structures in statistical research, and how frequently are they applied across domains?

(RQ2) In which domains are spatio-temporal models currently applied, and for what types of problems?

(RQ3) What are the key challenges and limitations in the application of spatio-temporal models across different domains?

We structure our work into several main sections. First, in Section \ref{sec:methodology}, we describe the methodology of our systematic literature review. Then, in Section \ref{sec:related_work}, we embed our work within the existing field of research. We present the results of our review in Section \ref{sec:results}, which forms the central part of the work. We detail the outcomes of the literature search in Section \ref{subsec:search-results}. Section \ref{subsec:modelling_techniques} provides an overview of the spatio-temporal model structures employed. These model structures are then considered in domain-specific contexts in Section \ref{subsec:application-domains}. Finally, Sections \ref{sec:discussion} and \ref{sec:conclusion} contain the discussion and conclusion, respectively.

All resources for our review are publicly available at \url{https://github.com/Media-Bias-Group/Spatio-Temporal-Statistical-Models-and-Their-Applications}.

%%%%%%%%%%%%%%%%%%%%%%%%%%%%%%%%%%%%%%%%%%%%%% Methodology
\section{Methodology}\label{sec:methodology}

We conduct a systematic review of published research that applies spatio-temporal statistical methods.

The primary contribution of our work is a comprehensive and structured synthesis of both the modeling frameworks and their applications across different domains. Literature reviews are often vulnerable to incomplete coverage and weaknesses in the selection, organization, and presentation of material \cite{fagan2017evidence}, particularly when comprehensive coverage is desired. To minimize these risks, we carefully designed our collection and selection procedures, focusing on consistency and transparency throughout the review process. We adhere to the Preferred Reporting Items for Systematic Reviews and Meta-Analyses (PRISMA) guidelines \cite{page2021prisma}, incorporating the following elements: Eligibility criteria (see Section \ref{subsec:eligibility-criteria}), information sources (see Section \ref{subsec:source_search-strategy}), search strategy (see Section \ref{subsec:source_search-strategy}), selection process (see Section \ref{subsec:selection_process}), data collection process (see Section \ref{subsec:data_extraction}), data items (see Section \ref{subsec:data_extraction}), and study risk of bias assessment (see Section \ref{subsec:QA}).

\subsection{Eligibility Criteria}\label{subsec:eligibility-criteria}

Our systematic literature review encompasses peer-reviewed journal articles and conference papers, including only studies published in English. We assume that current studies build upon knowledge accumulated in previous time periods. While many methods in the domain of spatio-temporal statistics originated many years ago (e.g., \cite{besag1991bayesian}), we aim to review methods used within the last few years, regardless of their initial development date. Therefore, we decided to cover the period from 2021 to 2025. We focus on research on spatio-temporal statistical modeling in applied domains. Theoretical works without application, as well as spatial-only or temporal-only approaches, were not included in the review. We also exclude pure mathematical work without a domain context or any work lacking a modeling focus. Additionally, we do not analyze publications in journals with a Scimago ranking lower than Q1\footnote{\url{https://www.scimagojr.com/journalrank.php} access 07.08.2025} or from conferences with a core rank lower than A\footnote{\url{https://portal.core.edu.au/conf-ranks/} access 07.08.2025}.

\subsection{Information Sources and Search Strategy}\label{subsec:source_search-strategy}

We queried the databases Scopus and Web of Science. Scopus covers a broader range of journals than Web of Science and offers faster citation analysis with greater article coverage \cite{fagan2017evidence}. However, Web of Science provides more detailed and visually informative citation analysis, reflecting its long-standing focus on meeting the needs of citation analysis research \cite{falagas2008comparison}. The corresponding search strings can be found in Table \ref{tab:searchString}. We applied the restriction to the \texttt{EXACTKEYWORD} terms ("Spatio-temporal Models" and "Spatiotemporal Analysis") for the Scopus search based on filtering options suggested by Scopus. Without this restriction, the search retrieved nearly twice as many records, most of which we would have excluded during the title screening stage. This additional constraint, therefore, enhances the precision of the search results and reduces the manual screening effort without substantially limiting the coverage of the research field. We conducted the query on 7 August 2025.

\subsection{Selection Process}\label{subsec:selection_process}

We imported every article retrieved from the databases described in Section \ref{subsec:source_search-strategy} to Zotero\footnote{\url{https://www.zotero.org/}} and removed duplicates. We then screened the titles and abstracts to identify eligible studies based on the exclusion criteria outlined in Section \ref{subsec:eligibility-criteria}. We made no exclusion if only one of several models described was a spatio-temporal statistical approach. Following the title and abstract screening, we conducted a full-text review. Studies that did not meet all quality assessment criteria (described in Section \ref{subsec:QA}) were excluded.

\subsection{Classification Scheme}
We choose a multi-step, qualitative approach to develop the classification scheme for statistical models (detailed in Section \ref{subsec:modelling_techniques} and illustrated in Figure \ref{fig:taxonomy}). First, we manually reviewed the included sources (detailed in Section \ref{subsec:search-results}). Then, we extracted the model structures described, comparatively analyzed them, and classified them according to their characteristics (i.e., statistical model type, spatial structure, temporal structure, spatio-temporal structure, and, if specified, assigned priors). The models were then grouped based on their structural similarities.

At the same time, existing classifications and systematizations from established textbooks \cite{cressie2011statistics, elhorst2014spatial, wikle2019spatio} were considered. We used both of these to validate the groups found and to develop and formulate the categories in terms of content. We developed the final structure of the scheme iteratively and made several adaptations to ensure both theoretical consistency and practical applicability.

\subsection{Data Collection Process and Data Items} \label{subsec:data_extraction}

For all studies included in the final collection, we extracted the following information to obtain a comprehensive overview of each study:

\textit{title, last name of the first author, year of publication, name of the journal, application area, data source, spatial data unit, temporal data unit, target variable, type of target variable, covariates, estimation technique, inference approach, software, availability of the code, statistical model type, spatial structure in the model, temporal structure in the model, spatio-temporal structure in the model, assigned priors, purpose of fitting a spatio-temporal model, challenges and limitations}. 

The \textit{assigned priors} category is required only for models that use Bayesian inference. When multiple models are described in a paper, we include the best model as specified by the authors' criteria. If the authors do not identify a single best model, we include all described models.

\subsection{Study Risk of Bias Assessment}\label{subsec:QA}

We assess the risk of bias for all included studies using the following quality assessment (QA) criteria.

(QA1) Does the paper explicitly define and focus on a statistical spatio-temporal model? 

(QA2) Is the model structure rigorously and transparently described in mathematical terms, and is it well-suited to the stated research question? 

(QA3) Is the application domain clearly specified? 

(QA4) Does the study explicitly apply the model to a spatio-temporal problem, and is this application central to the analysis? 

(QA5) Are the data types, variables, and data sources fully specified and described with sufficient detail? 

(QA6) Does the study include an appropriate model assessment? 

(QA7) Are the results accurately interpreted within the context of the study, including a clear discussion of limitations? 

The criteria are checked sequentially. We begin by evaluating criterion (QA1). If the paper meets (QA1), we proceed with the evaluation to the next criterion (QA2), and this process continues until all criteria have been evaluated. We included a paper only if all of the QA criteria are met. We documented the decision regarding which papers do not meet specific criteria in our repository available at \url{https://github.com/Media-Bias-Group/Spatio-Temporal-Statistical-Models-and-Their-Applications}.

%%%%%%%%%%%%%%%%%%%%%%%%%%%%%%%%%%%%%%%%%%%%%% Related Work
\section{Related Work}\label{sec:related_work}

\begin{figure}[h]
    \centering
    \begin{tikzpicture}

\fill (0,0) circle (0.1cm); 
\draw[->, thick] (0,0) -- (-5,-5);
\draw[->, thick] (0,0) -- (4.9,-4.9);
\draw[rounded corners, fill = gray, fill opacity=0.3] (-10, -10) rectangle (-5, -5); 
\node [align = center] at (-7.5,-7.5) {\huge \textbf{Data Mining} \\ \huge \textbf{Surveys} \\ \\ \huge (Section \ref{subsec:related_fields}) \\ \\ \Large \textcite{alam2022survey}, \\ \Large \textcite{koutsaki2023spatiotemporal}, \\ \Large \textcite{shekhar2015spatiotemporal}};
\fill (5,-5) circle (0.1cm); 

\draw[->, thick] (5,-5) -- (0.1,-9.9);
\draw[->, thick] (5,-5) -- (10,-10);
\draw[rounded corners, fill = gray, fill opacity=0.3] (10, -15) rectangle (15, -10); 
\node [align = center] at (12.5,-12.5) {\huge \textbf{ML Surveys }\\ \\ \huge (Section \ref{subsec:related_fields}) \\ \\ \Large \textcite{sun2024survey},\\ \Large \textcite{wang2020deep}};
\draw[rounded corners, fill = gray, fill opacity=0.3] (-2.5, -13) rectangle (2.5, -10); 
\node [align = center] at (0,-11.5) {\huge \textbf{Statistic}\\ \huge \textbf{Surveys}\\ \\ \huge (Section \ref{subsec:reviews})};

\draw[->, thick] (0,-13) -- (-5,-18);
\draw[->, thick] (0,-13) -- (4.9,-17.9);
\draw[rounded corners, fill = gray, fill opacity=0.3] (-10, -21) rectangle (-5, -18); 
\node [align = center] at (-7.5,-19.5) { \Large \textcite{gonzalez_spatio-temporal_2016},\\ \Large \textcite{reinhart_review_2018}};
\fill (5,-18) circle (0.1cm); 

\draw[->, thick] (5,-18) -- (-0.1,-22.9);
\draw[->, thick] (5,-18) -- (10,-23);
\draw[rounded corners, fill = gray, fill opacity=0.3] (-2.7, -26) rectangle (2.7, -23); 
\node [align = center] at (0,-24.5) { \Large \textcite{aswi_bayesian_2019},\\ \Large \textcite{byun_systematic_2021},\\ \Large \textcite{martenies_spatiotemporal_2021}, \\\Large \textcite{odhiambo_spatial_2023}, \\ \Large \textcite{tesema_systematic_2023}};
\draw[rounded corners, fill = black, fill opacity=0.8] (10, -26) rectangle (15, -23); 
\node [align = center, color = white] at (12.5,-24.5) { \huge \textbf{Our Review}};

\node [align = center, rotate = 45] at (-2.5,-2.5) { \huge no focus on \\ \\ \huge ST modeling};
\node [align = center, rotate = -45] at (2.5,-2.5) { \huge focus on \\ \\ \huge ST modeling};
\node [align = center, rotate = 45] at (2.5,-7.5) { \huge focus on \\ \\ \huge interp. structures};
\node [align = center, rotate = -45] at (7.5,-7.5) { \huge no focus on \\ \\ \huge interp. structures};
\node [align = center, rotate = 45] at (-2.5,-15.5) { \huge one specific \\ \\ \huge model class};
\node [align = center, rotate = -45] at (2.5,-15.5) { \huge overview of \\ \\ \huge model classes};

\node [align = center, rotate = 45] at (2.5,-20.5) { \huge one specific \\ \\ \huge appl. field};
\node [align = center, rotate = -45] at (7.5,-20.5) { \huge overview of \\ \\ \huge appl. fields};

\end{tikzpicture}
    \caption{
        Illustration of a classification of our contribution within the
        context of related survey and review papers.
        ST = Spatio-Temporal,
        ML = Machine Learning.
    }
    \label{fig:relatedWork}
\end{figure}
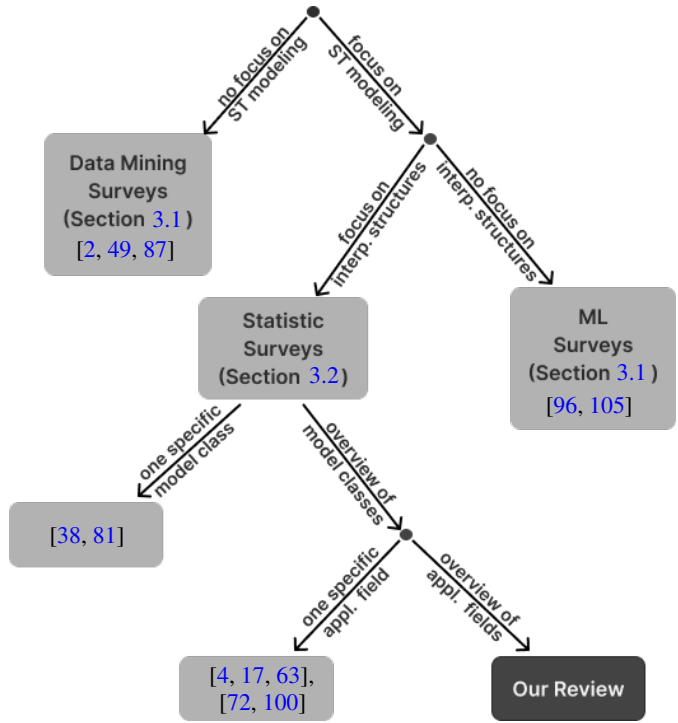

Figure \ref{fig:relatedWork} illustrates the gap that our review addresses relative to existing literature reviews. The following sections elaborate on the sources summarized in the figure: Section \ref{subsec:related_fields} includes surveys from related fields. Section \ref{subsec:reviews} comprises review papers on spatio-temporal statistics. Section \ref{subsec:textbooks} covers textbooks and theoretical works that provide an overview of foundational theories and model taxonomies.

\subsection{Work in Related Fields} \label{subsec:related_fields}
In our work, we focus on interpretable spatio-temporal modeling structures. Therefore, we distinguish our review from related fields with (i) no focus on spatio-temporal modeling or (ii) no interpretable modeling structures. 

Regarding (i), there are overviews of general spatio-temporal data mining approaches. Existing surveys (e.g., \cite{alam2022survey, koutsaki2023spatiotemporal, shekhar2015spatiotemporal}) already provide comprehensive insight into the theoretical foundations, data structures, and computer-aided methods for analyzing spatio-temporal data. They categorize existing systems and methods based on data types, computing architectures, and analytical objectives, considering databases, big data platforms, Geographic Information Systems (GIS), and statistical software \cite{alam2022survey, koutsaki2023spatiotemporal, shekhar2015spatiotemporal}. Key topics include clustering, outlier detection, and change analysis \cite{koutsaki2023spatiotemporal, shekhar2015spatiotemporal}. They place particular emphasis on statistical fundamentals such as spatial autocorrelation, non-stationarity, and tele-coupling \cite{shekhar2015spatiotemporal}. The works highlight applications in environmental research, epidemiology, climate science, crime analysis, and mobility research \cite{koutsaki2023spatiotemporal, shekhar2015spatiotemporal}. 

The second category (ii) refers to the field of machine learning (ML).
Spatio-temporal ML models are well-suited for predictive tasks. However, they reach their limits when interpretability and uncertainty quantification are critical, which means that we cannot make clear statements about which features affect the final results and how \cite{sun2024survey, wikle2023statistical}. 
Furthermore, comprehensive surveys already exist (e.g., \cite{sun2024survey, wang2020deep}) on spatio-temporal ML approaches. \cite{sun2024survey} systematically classified models for predicting spatio-temporal series using deep learning into a taxonomy. They discuss applications such as traffic flow and environmental monitoring, as well as key challenges, including modeling spatial dependencies, scalability, and interpretability. The authors demonstrate that classical multivariate time series ML models frequently overlook spatial correlations and network effects, and suggest that graph network techniques and transformer architectures should be given greater consideration. The paper \cite{wang2020deep} presents, next to a comprehensive overview of the use of deep learning techniques, a general pipeline framework that combines data representation, model selection, and tasks (e.g., prediction, anomaly detection, representation learning).

Another related field we do not cover is research on the fusion of deep learning with traditional statistical models, which enables more robust handling of uncertainty and enhances model transparency. Such approaches involve first constructing classical spatio-temporal statistical models and then integrating deep learning to characterize the conditional distributions \cite{wikle2023statistical}. 

\subsection{Review and Survey Papers}\label{subsec:reviews}

Our literature review provides an overview of different spatio-temporal model structures and discusses their applications in various domains. Therefore, we distinguish our work from (a) review papers, which give an overview of different application cases, but focus on one specific spatio-temporal model class, or (b) surveys, which provide an overview of different model classes, but focus on specific fields of application.

Examples for category (a) are \cite{gonzalez_spatio-temporal_2016} and \cite{reinhart_review_2018}, which focus on spatio-temporal point process models. Both works emphasize theoretical model structures while illustrating applications across diverse fields. \cite{gonzalez_spatio-temporal_2016} exposes point process methodology with applications ranging from Ebola outbreaks to tornadoes, while \cite{reinhart_review_2018} synthesizes self-exciting point process applications in earthquakes, criminology, and epidemics. Both \cite{gonzalez_spatio-temporal_2016} and \cite{reinhart_review_2018} provide detailed theoretical insights and broader interdisciplinary perspectives. However, they generally follow less formalized review procedures and tend to provide limited discussion of model comparison or evaluation criteria.

Examples for category (b) are the reviews \cite{aswi_bayesian_2019, byun_systematic_2021, martenies_spatiotemporal_2021, odhiambo_spatial_2023}, and \cite{tesema_systematic_2023}, which concentrate on specific application fields but differ in terms of thematic scope, methodological depth, and classification approach.
Among them, \cite{aswi_bayesian_2019} provides the most detailed methodological treatment, analyzing Bayesian spatial and spatio-temporal models for modeling dengue fever with a focus on priors and covariates. The authors of \cite{odhiambo_spatial_2023} also adopt a disease-specific approach for COVID-19, providing a descriptive overview of the applied models, with less emphasis on underlying assumptions or inferential frameworks. In contrast, \cite{byun_systematic_2021} expands the application scope to public health in Korea, integrating not only regression-based models but also clustering and interpolation techniques. This expansion, however, is accompanied by a more general treatment of methodological foundations. \cite{tesema_systematic_2023} narrows the focus to joint spatial and spatio-temporal models, emphasizing their advantages for multi-outcome health research, while placing less focus on model types beyond this subset. \cite{martenies_spatiotemporal_2021} discusses methodologies, particularly Bayesian hierarchical approaches, with a specific focus on species distribution modeling.

The authors of \cite{sahu_recent_nodate_2005} combine elements from both traditions, reviewing various model classes, including point processes, covariance models, and Bayesian filtering approaches, as well as applications in environmental monitoring, criminology, and archaeology. While methodologically diverse, the review does not follow a systematic protocol and only covers models until 2005.

\subsection{Textbooks and Overview Works}\label{subsec:textbooks}
In our work, we provide a classification scheme for spatio-temporal model structures. We focus our review on model structures that are currently in use in practice. Therefore, we distinguish our contribution from the taxonomies and classification schemes found in textbooks and theoretical overview works. Nevertheless, we mention that we used the works discussed below as a baseline for developing our classification scheme.

Publications such as \cite{cressie2011statistics, elhorst2014spatial, wikle2015modern}, and \cite{wikle2019spatio} establish the conceptual basis of spatio-temporal statistics, though they differ in scope, classification logic, and intended audience.
The theoretical overview paper \cite{wikle2015modern} provides a concise introduction, emphasizing the distinction between descriptive covariance-based models and dynamic conditional process models. While this classification provides analytical clarity, it does not encompass the full range of spatio-temporal modeling approaches, such as point processes. In comparison, the comprehensive volume by \cite{cressie2011statistics} presents an extensive methodological synthesis. Covering covariance models, stochastic differential equations, geostatistical and lattice processes, as well as point processes, it offers one of the broadest classifications available in textbook literature. Its emphasis on hierarchical dynamical models provides a coherent framework, though it places less attention on alternative modeling strategies and includes limited empirical illustrations. 

The authors of \cite{wikle2019spatio} adopt a different approach; instead of organizing models by data type or mechanics, they classify them according to analytic goals such as prediction, forecasting, and inference. This application-oriented framing supports practical implementation but provides less systematic coverage of methodological distinctions. The textbook \cite{elhorst2014spatial} adopts an econometric perspective, categorizing spatial econometric models into three types: Cross-sectional, panel, and dynamic panel data models. While clearly structured and methodologically rigorous, this typology is situated within the econometric tradition and focuses primarily on frequentist approaches, with less emphasis on Bayesian or hierarchical methods.

\subsection{Summary}
Consequently, the current literature presents a heterogeneous picture. Methodological classifications tend to reflect specific focal points \cite{elhorst_spatial_2022, wikle2015modern}, focused on hierarchical Bayes \cite{cressie2011statistics}, or organized around applied goals \cite{wikle2019spatio}. Review papers, on the other hand, provide either methodological rigor with a relatively confined scope \cite{aswi_bayesian_2019, byun_systematic_2021, odhiambo_spatial_2023, tesema_systematic_2023} or conceptual depth without employing a systematic review protocol \cite{gonzalez_spatio-temporal_2016, martenies_spatiotemporal_2021, reinhart_review_2018, sahu_recent_nodate_2005}. A systematic review that integrates classification schemes with comprehensive application contexts is still lacking. Additionally, a further limitation of the existing body of work is its temporal distribution. Some contributions, such as \cite{sahu_recent_nodate_2005}, were published before many now-established approaches, like integrated nested Laplace approximations (established in 2009 by \cite{rue2009approximate}), were available. This temporal gap highlights the need for updated syntheses that incorporate recent developments and provide a more contemporary classification of spatio-temporal modeling approaches. Addressing this gap constitutes the main contribution of our work.

%%%%%%%%%%%%%%%%%%%%%%%%%%%%%%%%%%%%%%%%%%%%%% Results
\section{Results}\label{sec:results}
This chapter presents the results of our systematic literature review. We begin by summarizing the outcomes of the search strategy outlined in Section \ref{sec:methodology} (see Section \ref{subsec:search-results}). Next, we analyze the content of the reviewed papers, focusing on the model structures employed (Section \ref{subsec:modelling_techniques}), and a detailed breakdown of the application domains (Section \ref{subsec:application-domains}).

\subsection{Results of the Search Strategy}\label{subsec:search-results}
\begin{figure}[h]
    \centering
    \includegraphics[width = 0.8\linewidth, trim={1.5cm 11.5cm 6.5cm 2.5cm}, clip]{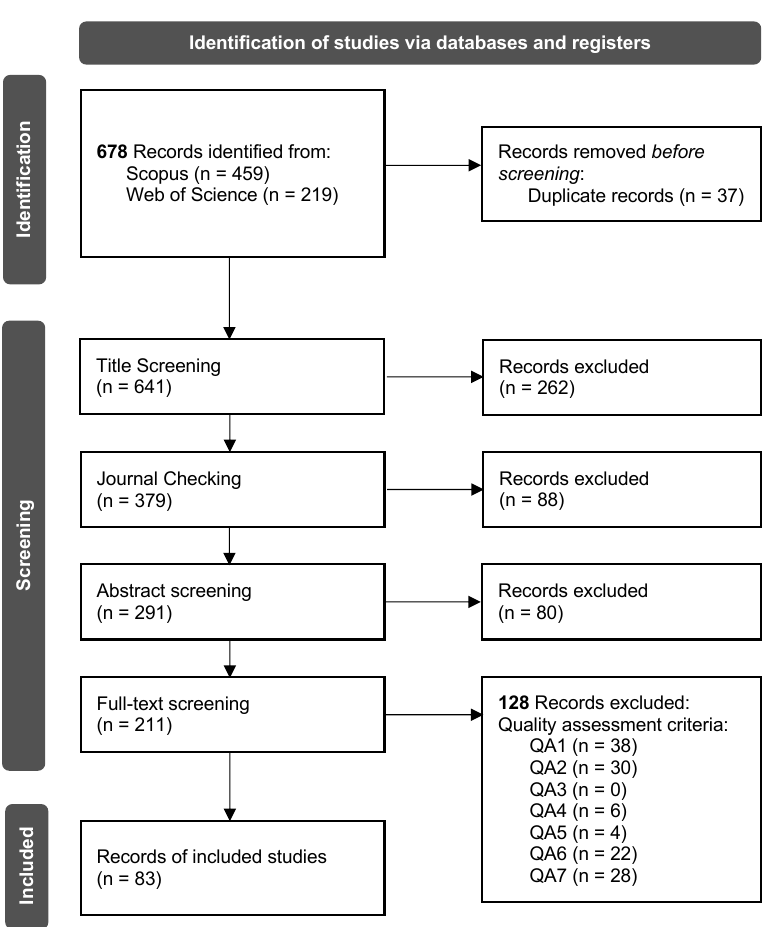}
    \caption{PRISMA flow diagram for new systematic reviews, which included searches of databases and registers only \cite{page2021prisma}.}
    \label{fig:PRISMA}
\end{figure}

We illustrate the search process in the PRISMA flow diagram shown in Figure \ref{fig:PRISMA} and documented in our repository (see URL in Section \ref{sec:introduction}). From the initial search, we identified a total of 678 studies: 459 from Scopus and 219 from Web of Science. After removing 37 duplicate records and excluding 262 papers after title screening, we checked the journal/conference ranking of the remaining 379 sources. This step resulted in the exclusion of 88 papers. We then screened the abstracts of the remaining 291 sources and excluded 80 based on this. After that, we conducted a full-text review, which led to the identification of 128 reports that did not meet all quality assurance (QA) criteria (see Section \ref{subsec:QA}). Finally, we included 83 publications in this review.

\subsection{Classification Scheme for Spatio-Temporal Modeling Structures}\label{subsec:modelling_techniques}
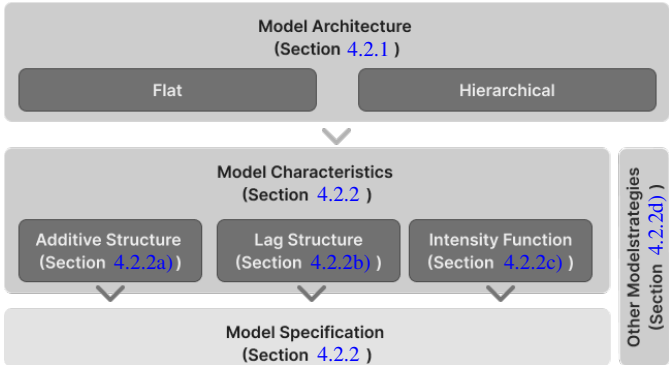
\begin{figure}[h]
    \centering
    \begin{tikzpicture}[transform shape]

\draw[rounded corners, fill = gray, fill opacity=0.3] (0, 0.5) rectangle (20, 3); 
\node [align = center] at (10,1.75) {\huge \textbf{Model Specification} \\ \\ \huge (Section \ref{subsub:L2-3}) };

\draw[rounded corners, fill = gray, fill opacity=0.3] (0, 3.5) rectangle (20, 9); 
\node [align = center] at (10,7.5) {\huge \textbf{Model Characteristics} \\ \\ \huge (Section \ref{subsub:L2-3}) };
\draw[rounded corners, fill = black, fill opacity=0.8] (0.5, 4) rectangle (6.5, 6); 
\node [align = center, color = white] at (3.5,5) {\LARGE \textbf{Additive Structure} \\ \\ \LARGE (Section \ref{subsec:ST-CoV}) };
\draw[rounded corners, fill = black, fill opacity=0.8] (7, 4) rectangle (13, 6); 
\node [align = center, color = white] at (10,5) {\LARGE \textbf{Lag Structure} \\ \\ \LARGE (Section \ref{subsec:lag-structures}) };
\draw[rounded corners, fill = black, fill opacity=0.8] (13.5, 4) rectangle (19.5, 6); 
\node [align = center, color = white] at (16.5,5) {\LARGE \textbf{Intensity Function} \\ \\ \LARGE (Section \ref{subsec:PP}) };

\draw[rounded corners, fill = gray, fill opacity=0.3] (20.5, 0.5) rectangle (24.5, 9); 
\node at (22.5,4.75) {
    \rotatebox{90}{
        \begin{tabular}{c}
            \huge \textbf{Other Modelstrategies} \\ \\ 
            \huge (Section \ref{subsub:other-forms})
        \end{tabular}
    }
};

\draw[rounded corners, fill = gray, fill opacity=0.3] (0, 9.5) rectangle (24.5, 14.5); 
\node [align = center] at (12.25,13) {\huge \textbf{Model Architecture} \\ \\ \huge (Section \ref{subsub:L1}) };
\draw[rounded corners, fill = black, fill opacity=0.8] (0.5, 10) rectangle (11.75, 11.5); 
\node [align = center, color = white] at (5.875,10.75) {\LARGE \textbf{Flat}};
\draw[rounded corners, fill = black, fill opacity=0.8] (12.75, 10) rectangle (24, 11.5); 
\node [align = center, color = white] at (18.5,10.75) {\LARGE \textbf{Hierarchical}};

\node[isosceles triangle, isosceles triangle apex angle=90, draw, fill=gray, minimum size =0.8cm, rotate = -90] (T90)at (12.25,9.5){};
\node[isosceles triangle, isosceles triangle apex angle=90, draw, fill=gray, minimum size =0.8cm, rotate = -90] (T90)at (3.5,3.5){};

\node[isosceles triangle, isosceles triangle apex angle=90, draw, fill=gray, minimum size =0.8cm, rotate = -90] (T90)at (10,3.5){};

\node[isosceles triangle, isosceles triangle apex angle=90, draw, fill=gray, minimum size =0.8cm, rotate = -90] (T90)at (16.5,3.5){}; 

\end{tikzpicture}
    \caption{Illustration of our proposed scheme for classifying model structures for spatio-temporal data.}
    \label{fig:taxonomy}
\end{figure}

The goal of this section is to provide an overview of the spatio-temporal statistical model structures used in the reviewed literature, as detailed in Section \ref{subsec:search-results}. To achieve this, we propose a classification scheme for spatio-temporal model structures, which is illustrated in Figure \ref{fig:taxonomy}. The classification of all reviewed models is presented in Table \ref{tab:classification-models}, and we provide a detailed description of each category below. 

To describe the individual levels of the classification scheme at a moderate mathematical level, we first introduce the following notation. Let $D_s$ be a spatial index set, i.e., a collection of spatial locations (e.g., geographical coordinates), and let $D_t$ be a set of temporal indices. Their product $D_s \times D_t$ forms a spatio-temporal index set. Define the latent or true process as $Y := \{Y_{t,s} \mid (t,s) \in D_t \times D_t\}$ and the observed data process as $Z := \{Z_{t,s} \mid (t,s) \in D_t \times D_t\}$. Furthermore, for a random variable $A$, we denote its probability distribution by $[A]$.

\subsubsection{Model Architecture.} \label{subsub:L1}
At the first level of our classification scheme, we distinguish between two main model architectures: Flat models and hierarchical models, e.g., as discussed in \cite{cressie2011statistics}.

\textit{Flat Model Architecture.} A flat model architecture only models the data process $Z$. For example, we can represent a flat spatio-temporal model as $Z_{t,s} = X_{t,s}^T\beta + \psi_{t,s} + \epsilon_{t,s}$, where $X_{t,s}^T\beta$ is the linear predictor, $\psi_{t,s}$ is the spatio-temporal component and $\epsilon_{t,s}$ is the error term for time $t$ and spatial unit $s$. Inference is based directly on the distribution of the data, either via likelihood-based estimation of $[Z \mid \theta]$ in a frequentist setting, or posterior inference $[\theta \mid Z]$ in a Bayesian setting, where $\theta$ denotes the model parameter. However, this flat specification does not explicitly model the latent process $Y$. 

\textit{Hierarchical Model Architecture.} Hierarchical models overcome this limitation by factorizing the joint distribution into conditional distributions \cite{wikle2003hierarchical}. These models explicitly distinguish between the latent process $Y$, which represents the underlying spatio-temporal phenomenon of interest, and the observed data process $Z$. Since empirical data are often incomplete and subject to measurement error, we treat the observed data as realizations of the observed data process $Z$, which is conditionally linked to $Y$ through a data model. Formally, we express this as $[Z \mid Y, \theta_D]$, where $\theta_D$ are the parameters of the data model. The latent process itself is described by a process model, $[Y \mid \theta_P]$, with $\theta_P$ denoting the process parameters. This results in the following hierarchical model structure:  

Data model: $[Z \mid Y, \theta_D]$

Process model: $[Y \mid \theta_P]$

\noindent The joint likelihood is then given by $[Y \mid Z, \theta_D, \theta_P] \propto [Z \mid Y, \theta_D]\,[Y \mid \theta_P]$.  An example of a hierarchical model formulation is the zero-inflated Poisson distribution model of \cite{jeong_investigating_2023}:

Data model: $Z_{t,s}\vert Y_{t,s}, \theta_D  \sim \textit{Poisson}(\theta_D(1-Y_{t,s}))$

Process model: $Y_{t,s}\vert \theta_P \sim \textit{Bernoulli}(\theta_P)$

\noindent In the application \cite{jeong_investigating_2023}, the data model describes the observed count of disease cases, depending on the Poisson mean $\theta_D$ and whether a structural zero occurs (i.e., the latent process $Y$). 

In a Bayesian framework, prior distributions are introduced for the parameters $\theta_D$ and $\theta_P$, leading to the posterior distribution $[Y, \theta_D, \theta_P \mid Z] \propto [Z \mid Y, \theta_D]\,[Y \mid \theta_P]\,[\theta_D, \theta_P]$. The advantage of hierarchical models becomes apparent when model dependencies grow complex. By splitting the model into multiple levels, each component can be further structured (e.g., incorporating random effects).

\subsubsection{Model Characteristics and Specification.}\label{subsub:L2-3}
Model characteristic describes the components that capture the temporal and spatial structure, as well as the mathematical operations used. Model specification, in contrast, refers to the detailed definition of these components.
Since model characteristics and specification are closely related, we discuss both together in this section.

\paragraph{Spatio-Temporal Additive Structure.}\label{subsec:ST-CoV}
The majority of models in the reviewed literature incorporate spatio-temporal additive components.

\textit{Model Characteristic.} Models with a spatio-temporal additive structure can be expressed in the simplified form
\begin{equation*}
    Y_{t,s} = X_{t,s}^T\beta + \psi_{t,s}.
\end{equation*}
We can consider $Y$ either as a Gaussian process ($\mathcal{GP}$) or a Markov random field (MRF). However, to specify realistic spatio-temporal dependence, we often separate the spatio-temporal effect $\psi_{t,s}$ into additive spatial, temporal, and spatio-temporal components, denoted by $u_s$, $v_t$, and $\gamma_{t,s}$, respectively. As summarized in Table \ref{tab:decomp-STeffect}, various decompositions of the additive component $\psi_{t,s}$ appear in the literature.
\begin{table}[h]
\footnotesize
\begin{tabularx}{\linewidth}{lX}
\hline
Component $\psi_{t,s}$  & Reference \\ \hline 
$u_s + v_t + \gamma_{t,s}$         &   \cite{bofa_optimizing_2024, briz-redon_impact_2021, briz-redon_comparison_2022, chen_spatio-temporal_2023, chirombo_prevalence_2024, ibeji_bayesian_2022, li_spatio-temporal_2023, liu_updated_2021, lome-hurtado_patterns_2021, mahfoud_forecasting_2021, ngwira_spatial_2021, orozco-acosta_scalable_2023, pasanen_spatio-temporal_2024, petrof_using_2024, satorra_bayesian_2024, vicente_multivariate_2023, wang_spatio-temporal_2025, zaouche_bayesian_2023}\\

$u_s + \gamma_{t,s}$  & \cite{fioravanti_spatio-temporal_2021, freitas_spatio-temporal_2021, griffith_spatial-temporal_2021, gruss_integrating_2023, iddrisu_spatio-temporal_2022, lindmark_evaluating_2023, nnanatu_evaluating_2021, olmos_estimating_2023, racek_capturing_2025, tam_bayesian_2024, yip_spatio-temporal_2022, zhang_high-resolution_2024, zhang_bayesian_2025} \\

$v_t + \gamma_{t,s}$  & \cite{epstein_mapping_2023, staeudle_accounting_2024}                  \\

$u_s + v_t$  & \cite{baldoni_agricultural_2021, barcelo_spatiotemporal_2024, briz-redon_bayesian_2024, briz-redon_association_2022, carroll_community_2021, dambrine_characterising_2021, elhorst_spatial_2022, freshwater_integrated_2021, gracia_chronic_2021, he_identifying_2021, huang_driving_2021, jalilian_hierarchical_2021, jeong_investigating_2023, ma_bayesian_2023, maia_migration_2021, pietak_effect_2024, pirani_effects_2024, sun_spatio-temporal_2021, yan_how_2025, zhao_dynamic_2022} \\

$\gamma_{t,s}$  & \cite{beloconi_spatio-temporal_2021, bofa_bayesian_2024, neupane_novel_2022, yin_bayesian_2024}\\ \hline
\end{tabularx}
\caption{Possible decomposition of the spatio-temporal effect $\psi_{s,t}$.}\label{tab:decomp-STeffect}
\end{table}
The spatio-temporal additive structure model may not be identifiable without substantial prior knowledge about the distributional properties of its components. We can model the terms $u_s$, $v_t$, and $\gamma_{t,s}$ using various formulations that capture spatial, temporal, or joint spatio-temporal dependence. 

\textit{Model Specification - Spatial Component.} 
Table \ref{tab:spatialEffect-structures} summarizes the spatial component specifications we find in the reviewed publications.

For the spatial component $u_s$, classical MRF formulations include the conditional auto-regressive (CAR) and intrinsic CAR (iCAR) models, which define local dependence between neighboring areas (denoted by $s^\prime \sim s$):
\begin{equation*}
    u_s \mid u_{-s} \sim \mathcal{N}\left( \frac{\sum_{s^\prime \sim s} w_{ss^\prime} u_{s^\prime}}{\sum_{s^\prime \sim s} w_{ss^\prime}}, \frac{\sigma_u^2}{\sum_{s^\prime \sim s} w_{ss^\prime}} \right),
\end{equation*}
where $w_{ss^\prime}$ denotes the spatial weight between units $s$ and $s^\prime$. Also, spatial random walks (RW), defined through neighbor differences $ u_s-u_j \sim \mathcal{N}(0,\sigma^2)$, are generally MRF.

The Besag–York–Molli\'e (BYM) model \cite{besag1991bayesian} decomposes the spatial effect into structured and unstructured components: $u_s = u_s^{(1)} + u_s^{(2)}$, where $u_s^{(1)} \sim \text{CAR}$ and $u_s^{(2)} \sim \mathcal{N}(0, \tau^{2})$. Its reparameterized version, BYM2, is given by $u_s = \sqrt{{(1-\rho)}/{\phi}}\, u_s^{(2)} + \sqrt{{\rho}/{\phi}}\, u_s^{(1)}$, where $\phi$ is the precision parameter and $\rho \in [0,1]$ weights the variability between $u_s^{(1)}$ and $u_s^{(2)}$. The BYM2 improves identifiability by scaling the structured and unstructured components. 

Gaussian processes ($\mathcal{GP}$) are specified through a covariance function $C_S(d_{s,s^\prime})$, where $d_{ss'}$ is the some distance between spatial units $s$ and $s^\prime$. For instance, the Mat\'ern covariance function, which is given through $C_{\text{Mat\'ern}}(d_{s,s^\prime}) = \sigma^2 ({2^{1-\nu}}/{\Gamma(\nu)}) (\kappa d_{ss'})^\nu K_\nu(\kappa d_{ss'})$ where $d_{ss'}$ is the euclidean distance, $\nu$ controls smoothness, $\kappa$ controls the spatial scale and $K_\nu$ denotes the modified Bessel-function of second kind. Simpler alternatives are independent Gaussian random effects (IID), $u_s \sim \mathcal{N}(0, \sigma_u^2)$.

Alternative modeling techniques for capturing spatial dependencies or effects include spline-based smoothers, given by $u(s) = \sum_{k=1}^K b_k(s)\beta_k$, which approximate nonlinear spatial trends through penalized basis expansions, and fixed effects (FE), which correspond to discrete region-specific parameters $u_s$. 
\begin{table}[h]
\footnotesize
\begin{tabularx}{\linewidth}{lX}
\hline
\textbf{Structure} & \textbf{References}  \\
\hline
BYM/BYM2 & \cite{briz-redon_impact_2021, briz-redon_bayesian_2024, briz-redon_comparison_2022, briz-redon_association_2022, chen_spatio-temporal_2023, gracia_chronic_2021, ibeji_bayesian_2022, iddrisu_spatio-temporal_2022, jalilian_hierarchical_2021, jeong_investigating_2023, lome-hurtado_patterns_2021, ma_bayesian_2023, mahfoud_forecasting_2021, ngwira_spatial_2021, orozco-acosta_scalable_2023, petrof_using_2024, satorra_bayesian_2024, yip_spatio-temporal_2022}\\

FE & \cite{baldoni_agricultural_2021, costantino_spatial_2023, elhorst_spatial_2022, freshwater_integrated_2021, he_identifying_2021, huang_driving_2021, li_spatio-temporal_2023, lozano_spatio-temporal_2023, maia_migration_2021, pietak_effect_2024, yan_how_2025, zhao_dynamic_2022}\\

IID & \cite{barcelo_spatiotemporal_2024, carroll_community_2021, fioravanti_spatio-temporal_2021, griffith_spatial-temporal_2021, li_spatio-temporal_2023, liu_updated_2021, nnanatu_evaluating_2021, pasanen_spatio-temporal_2024, sun_spatio-temporal_2021, zaouche_bayesian_2023, zhang_high-resolution_2024}\\

$\mathcal{GP}(0, C_S(d_{s,s^\prime}))$  & \cite{barcelo_spatiotemporal_2024, dambrine_characterising_2021, gruss_integrating_2023, lindmark_evaluating_2023, olmos_estimating_2023, saez_spatial_2022, wang_spatio-temporal_2025}  \\

CAR/iCAR & \cite{bofa_optimizing_2024, chirombo_prevalence_2024, freitas_spatio-temporal_2021, pirani_effects_2024, tam_bayesian_2024, zhang_bayesian_2025}\\

Smoothing function & \cite{racek_capturing_2025, vicente_multivariate_2023}\\

GMRF & \cite{jalilian_hierarchical_2021}\\ 

Spatial RW & \cite{nnanatu_evaluating_2021} \\

Moran eigenvector Filtering & \cite{griffith_spatial-temporal_2021} \\
\hline
\end{tabularx}
\caption{Model structures for the spatial effect $u_s$. GMRF = Gaussian Markov random field}\label{tab:spatialEffect-structures}
\end{table}

\textit{Model Specification - Temporal Component.} Table \ref{tab:temporalEffect-structures} summarizes the temporal component specifications we find in the reviewed publications. We can model the temporal component $v_t$ similarly to the spatial component. Standard specifications include a temporal CAR, auto-regressive (AR) models of order $p$, defined as $v_t = \sum_{k=1}^p \rho_k v_{t-k} + \varepsilon_t$, or random walks (RW) of order $p$ defined by $\Delta^p v_t \sim \mathcal{N}(0,\sigma_v^2)$. 

Additionally, temporal splines of the form $v(t) = \sum_{k=1}^K b_k(t)\beta_k$
allow flexible smoothing of long-term trends. Simpler alternatives include independent and identically distributed (IID) Gaussian noise or fixed effects (FE) as baseline specifications. 
\begin{table}[h]
\footnotesize
\begin{tabularx}{\linewidth}{lX}
\hline
\textbf{Structure} & \textbf{References}  \\
\hline
RW(p) & \cite{barcelo_spatiotemporal_2024, briz-redon_impact_2021, briz-redon_bayesian_2024, briz-redon_comparison_2022, carroll_community_2021, freshwater_integrated_2021, ibeji_bayesian_2022, jalilian_hierarchical_2021, jeong_investigating_2023, mahfoud_forecasting_2021, ngwira_spatial_2021, orozco-acosta_scalable_2023, petrof_using_2024, pirani_effects_2024, satorra_bayesian_2024, staeudle_accounting_2024, sun_spatio-temporal_2021} \\

IID &  \cite{briz-redon_impact_2021, briz-redon_bayesian_2024, briz-redon_comparison_2022, chen_spatio-temporal_2023, dambrine_characterising_2021, jeong_investigating_2023, lindmark_evaluating_2023, liu_updated_2021, lome-hurtado_patterns_2021, ma_bayesian_2023, mahfoud_forecasting_2021, ngwira_spatial_2021, pasanen_spatio-temporal_2024, tam_bayesian_2024}\\

FE & \cite{elhorst_spatial_2022, gracia_chronic_2021, he_identifying_2021, huang_driving_2021, maia_migration_2021, pietak_effect_2024, yan_how_2025, zhao_dynamic_2022}\\

AR(p) & \cite{chen_spatio-temporal_2023, jalilian_hierarchical_2021, lindmark_evaluating_2023, ma_bayesian_2023, wang_spatio-temporal_2025, zaouche_bayesian_2023}\\

Smoothing function & \cite{briz-redon_association_2022, epstein_mapping_2023, freshwater_integrated_2021, martenies_spatiotemporal_2021, vicente_multivariate_2023} \\

Temporal CAR & \cite{bofa_optimizing_2024, chirombo_prevalence_2024, zhang_bayesian_2025}\\
\hline
\end{tabularx}
\caption{Model structures for the temporal effect $v_t$.}\label{tab:temporalEffect-structures}
\end{table}

\textit{Model Specification - Spatio-Temporal Component.} Table \ref{tab:spatio-temporalEffect-structures} summarizes the specifications we find in the reviewed publications.

Spatio-temporal dependence, captured by $\gamma_{t,s}$, can be modeled using separable or non-separable covariance structures in $\mathcal{GP}$s. In the separable case, we specify the covariance function as $C_{ST}((s,t),(s',t')) = C_S(d_{ss'}) \cdot C_T(\vert t-t'\vert)$, where $C_S$ and $C_T$ are spatial and temporal covariance functions, respectively (see, e.g., \cite{knorr2000bayesian} for details). More general non-separable covariance functions include the Gneiting class. 

Dynamic formulations combine temporal auto-regression with spatial Gaussian processes. For example, $\gamma_{t,s} = \rho \gamma_{t-1,s} + c_{t,s}$, with the innovation term $c_{t,s}$ satisfying 
\begin{equation*}
    \text{Cov}\big(c(t, s), c(t', s^\prime)\big) =
\begin{cases}
0, & \text{if } t \neq t' \\
C(d_{ss'}), & \text{if } t = t'
\end{cases}.
\end{equation*}
Additionally, tensor product splines of the form $w(t,s) = \sum_{k=1}^K \sum_{\ell=1}^L b_k(s)c_\ell(t)\beta_{k\ell}$,
offer a flexible, nonparametric approach for representing smooth interactions between space and time.
\begin{table}[h]
\footnotesize
\begin{tabularx}{\linewidth}{lX}
\hline
\textbf{Structure} & \textbf{References}  \\
\hline
Separable covariance function & \cite{bofa_optimizing_2024, briz-redon_impact_2021, briz-redon_comparison_2022, chen_spatio-temporal_2023, chirombo_prevalence_2024, liu_updated_2021, lome-hurtado_patterns_2021, mahfoud_forecasting_2021, ngwira_spatial_2021, nnanatu_evaluating_2021, olmos_estimating_2023, orozco-acosta_scalable_2023, petrof_using_2024, satorra_bayesian_2024, tam_bayesian_2024,  vicente_multivariate_2023, wang_spatio-temporal_2025, yip_spatio-temporal_2022}\\

AR with spatial corr. innovation & \cite{beloconi_spatio-temporal_2021, bofa_bayesian_2024, fioravanti_spatio-temporal_2021, freitas_spatio-temporal_2021, gruss_integrating_2023, li_spatio-temporal_2023, staeudle_accounting_2024, yin_bayesian_2024, zaouche_bayesian_2023} \\

Smoothing function & \cite{epstein_mapping_2023, ibeji_bayesian_2022, racek_capturing_2025, zhang_high-resolution_2024}\\

Moran eigenvector filtering & \cite{griffith_spatial-temporal_2021} \\

Spatio-temporal slope & \cite{iddrisu_spatio-temporal_2022} \\

Gneiting covariance function & \cite{neupane_novel_2022}\\
\hline
\end{tabularx}
\caption{Model structures for the spatio-temporal effect $\gamma_{t,s}$.}\label{tab:spatio-temporalEffect-structures}
\end{table}

Together, these formulations highlight the wide range of possibilities for structuring spatio, temporal or spatio-temporal effects, from highly structured dependencies (CAR, Mat\'ern covariance function) to flexible smoothers (splines) and simpler baselines (IID or FE).

\paragraph{Lag Structures.}\label{subsec:lag-structures}
Another common approach for modeling spatio-temporal dependencies is the use of lag structures. For simplicity, we define the response vector as $Y_t:= [Y_{t,s_1}, \dots, Y_{t,s_N}]^T$ for $t \in D_t := \{0,1,2, \dots\}$ and $s_1,\dots,s_N \in D_s$ for the remainder of this section.

\textit{Model Characteristics.} 
Lag structures introduce dependencies across time or space by incorporating past values or values from neighboring units into the model. We distinguish between spatial lags, temporal lags, and spatio-temporal lags, the latter being a combination of the first two. Below, we examine each in detail.

\textit{Model Specification - Spatial Lag Structures.}
We typically introduce spatial dependencies via a spatial weight matrix
\begin{equation*}
    W = \begin{bmatrix}
        w_{s_1,s_1} & \dots & w_{s_1,s_N} \\
        \vdots & \ddots & \vdots \\
        w_{s_N,s_1} & \dots & w_{s_N,s_N}
    \end{bmatrix}.
\end{equation*}
This $N\times N$ matrix can take various forms, depending on the specific application. Common examples include: Binary neighborhood matrices, where $w_{s_i,s_j} = 1$ if $s_i$ and $s_j$ are neighbors, $w_{s_i,s_j} = 0$ otherwise (e.g., \cite{huang_driving_2021, lozano_spatio-temporal_2023, tam_bayesian_2024}); K-nearest neighbors (KNN) matrices, where $w_{s_i,s_j} = 1$ if $s_j$ is one of the KNN of $s_i$, $w_{s_i,s_j} = 0$ otherwise (e.g., \cite{costantino_spatial_2023}); inverse distance matrices, with $w_{s_i,s_j} = {1}/{d_{s_i,s_j}}$, where $d_{s_i,s_j} > 0$ is often based on the physical or geographical distance between locations $s_i$ and $s_j$ (e.g., \cite{costantino_spatial_2023}).
The matrix is often row-standardized: $w_{s_i,s_j}^{*} = w_{s_i,s_j} / \sum_j w_{s_i,s_j}$ (e.g., \cite{baldoni_agricultural_2021, maia_migration_2021}).

Different types of spatial lag structures correspond to various sources of spatial association: Spatial error for clustering in unobservables (see Equation \eqref{eq:1}), spatially lagged covariates for exogenous spillovers (see Equation \eqref{eq:2}), and spatial-lag/spatial-auto-regressive structures for the target variable (see Equation \eqref{eq:3}). We can write in simplified form:
\begin{eqnarray}
    Y_t &=& X_t\beta + u_t, \; \text{where} \; u_t = \lambda W u_t + \epsilon_t  \label{eq:1}\\
    Y_t &=& X_t\beta + WX_t \theta + \epsilon_t \label{eq:2}\\
    Y_t &=& \rho W Y_t + X_t\beta + \epsilon_t \label{eq:3}
\end{eqnarray}

\textit{Model Specification - Temporal Lag Structures.}
Temporal lag structures mirror their spatial counterparts and also represent different sources of temporal association: Time lag in the error term see Equation \eqref{eq:1-T}), temporal lagged covariates (see Equation \eqref{eq:2-T}), or temporal lagged/ auto-regressive target variable (see Equation \eqref{eq:3-T}). Temporal weights, denoted by $\Tilde{w}_l$ for lag $l = 1, \dots, p$ with $p < t \in D_t$, may also be defined. We can express the models as:
\begin{eqnarray}
    Y_t &=& X_t\beta + u_t, \; \text{where} \; u_t = \delta \sum_{l = 1}^{p} \Tilde{w}_l u_{t-l} + \epsilon_t  \label{eq:1-T}\\
    Y_t &=& X_t\beta + \gamma\sum_{l = 1}^{p} \Tilde{w}_l X_{t-l}  + \epsilon_t \label{eq:2-T}\\
    Y_t &=& \Phi\sum_{l = 1}^{p} \Tilde{w}_l  Y_{t-l} + X_t\beta + \epsilon_t \label{eq:3-T}
\end{eqnarray}

As noted, spatio-temporal lag structures arise from combinations of spatial and temporal lag specifications.

\textit{Model Specification - Other forms.}
We also find lag structures in conditional intensity, e.g., in \cite{yip_spatio-temporal_2022}, where the simplified model is given by
$Y_{t,s} \sim \text{NegBin}(e_{t,s}\lambda_{t,s}, r)$ with $\log(\lambda_{t,s})= a \log(\lambda_{t-1,s}) + X_{t,s}\beta$. Averaged or aggregated components (e.g., in \cite{beloconi_spatio-temporal_2021, baldoni_agricultural_2021, costantino_spatial_2023}) also contain indirect lag structures. Lastly, structures such as CAR, iCAR, AR, and RW, discussed earlier in Section \ref{subsub:L2-3} \ref{subsec:ST-CoV}, also fall into the category of lag structures. Since we already covered these under additive structures, we only briefly mention them here.

Table \ref{tab:Lag-structures} provides a summary of the lag structures found in the reviewed papers. 
\begin{table}[h]
\footnotesize
\begin{tabularx}{\linewidth}{lX}
\hline
\textbf{Structure} & \textbf{References}  \\
\hline
Spatial lag & \\
\hspace{0.1cm} error term & \cite{huang_driving_2021, maia_migration_2021, pietak_effect_2024, tepe_history_2024, yan_how_2025}\\
\hspace{0.1cm} covariate & \cite{barcelo_spatiotemporal_2024, pietak_effect_2024, tepe_history_2024, zhao_dynamic_2022}\\
\hspace{0.1cm} target variable & \cite{baldoni_agricultural_2021, costantino_spatial_2023, elhorst_spatial_2022, lozano_spatio-temporal_2023, maia_migration_2021, tepe_history_2024, yan_how_2025}\\
& \\
Temporal lag & \\
\hspace{0.1cm} covariate & \cite{baldoni_agricultural_2021, briz-redon_association_2022, lozano_spatio-temporal_2023, yan_how_2025, yip_spatio-temporal_2022, zhang_bayesian_2025} \\
\hspace{0.1cm} target variable & \cite{baldoni_agricultural_2021, costantino_spatial_2023, elhorst_spatial_2022, he_identifying_2021, lozano_spatio-temporal_2023, mattera_forecasting_2025, nguyen_impact_2023, tepe_history_2024, zhao_dynamic_2022}\\
& \\
Spatio-temporal lag & \\
\hspace{0.1cm} covariate & \cite{yan_how_2025}\\
\hspace{0.1cm} target variable & \cite{bracher_endemic-epidemic_2022, costantino_spatial_2023, elhorst_spatial_2022, nguyen_impact_2023, zhang_inferring_2022, zhao_dynamic_2022} \\
& \\
Other forms & \cite{baldoni_agricultural_2021, barcelo_spatiotemporal_2024, beloconi_spatio-temporal_2021, costantino_spatial_2023, gracia_chronic_2021, he_identifying_2021, lopez-lacort_bayesian_2024, yip_spatio-temporal_2022, zhang_inferring_2022}\\
\hline
\end{tabularx}
\caption{Lag structures in the models used in the reviewed literature.}\label{tab:Lag-structures}
\end{table}

\paragraph{Intensity Functions in Spatio-temporal Point Processes.}\label{subsec:PP}
Spatio-temporal point processes are stochastic models used to describe events occurring within a time-space domain $\mathbb{R} \times \mathbb{R}^d$, where each event is represented by a point $(t, s)$ in time and space. The event locations $(t, s)$ are not predetermined but are realized randomly. The process itself is a stochastic mechanism that generates random configurations of such points.

A central concept in these models is the intensity function $\lambda(t, s)$, which describes the rate or likelihood of events occurring within a small time-space region. The number of events $N$ within a region $D_t \times D_s$ is modeled using a general counting distribution $\mathcal{P}(D_t \times D_s)$, with the expected number of events given by the integral of the intensity function over that region: $\mathbb{E}[N(D_t \times D_s)] = \int_{D_t \times D_s} \lambda(t,s) \, dt \, ds,$. 
In some cases, we use the conditional intensity $\lambda(t,s \mid \mathcal{H})$, where $\mathcal{H}$ represents the history of past events. This conditional formulation captures the likelihood of an event occurring near $(t, s)$, given the occurrence of prior events (e.g., \cite{bei_predicting_2021, vomfell_no_2023}).

A more detailed treatment of spatio-temporal point processes and their model specifications can be found in \cite{gonzalez_spatio-temporal_2016, reinhart_review_2018}. Among the reviewed publications, we identify four applications of spatio-temporal point process models \cite{amaral_spatio-temporal_2023, bei_predicting_2021, valente_pre-harvest_2021, vomfell_no_2023}.

\paragraph{Other Spatio-Temporal Modeling Strategies.}\label{subsub:other-forms}
Table \ref{tab:spatio-temporalAlternatives} provides an overview of alternative modeling strategies proposed to capture complex spatio-temporal dynamics beyond classical additive or lag-based models. These approaches include domain-adapted state-space models, which extend conventional state-space formulations to incorporate system-specific constraints or observational features \cite{choi_short-term_2021, choi2025spatiotemporal, mcdonald_integrating_2023}. Stochastic compartment models, commonly used in epidemiology, represent interactions between latent compartments through probabilistic transitions \cite{liu_role_2025, nguyen_modelling_2023}. Another strategy is to model spatial and temporal interaction through the estimation method itself, as in geographically and temporally weighted regression (GTWR) \cite{chen_bayesian_2024, liu_influence_2023}. In addition, spatio-temporal extreme value models \cite{sando_multivariate_2024, garcia_bayesian_2023} and the Bayesian Maximum Entropy (BME) framework \cite{gomez_integrating_2021} provide flexible tools for accounting for tail behavior, local uncertainty, and complex dependency structures. These models are particularly well-suited for accommodating non-standard dynamics, strong nonlinearities, and interactions that may not be adequately captured by traditional additive or lag-structured spatio-temporal formulations.
\begin{table}[h]
\footnotesize
\begin{tabularx}{\linewidth}{lX}
\hline
\textbf{Structure} & \textbf{References}  \\
\hline
Domain-adapted state-space models & \cite{choi_short-term_2021, choi2025spatiotemporal, mcdonald_integrating_2023}\\
Stochastic compartment models & \cite{liu_role_2025, nguyen_modelling_2023}\\
GTWR &  \cite{chen_bayesian_2024, liu_influence_2023}\\
Spatio-temporal extreme value model &  \cite{garcia_bayesian_2023, sando_multivariate_2024}\\
spatially correlated self-exciting model & \cite{clark_class_2021} \\
BME &  \cite{gomez_integrating_2021}\\
\hline
\end{tabularx}
\caption{Alternative spatio-temporal model structures.}\label{tab:spatio-temporalAlternatives}
\end{table}

\subsection{Application domains}\label{subsec:application-domains}
Table \ref{tab:appl.areas} provides an overview of the application area distribution of the included studies. We identified six distinct fields of application in the reviewed publications and manually assigned each paper to one of the following categories: Epidemiology, ecology, public health, economics, and criminology. We cannot clearly assign five publications to any of these major fields, and we therefore categorize them as \textit{others}. We give a detailed description of each application domain in the Appendix \ref{apx:application-domains}. In the following, we summarize the main findings and combine them with the classifications described in Section \ref{subsec:modelling_techniques}.  
\begin{table}[h]
\footnotesize
\begin{tabularx}{\linewidth}{p{2.2cm}cX}
\hline
Application Domain & Frequency  & References \\ \hline
  Epidemiology     &   36       &  \cite{amaral_spatio-temporal_2023, barcelo_spatiotemporal_2024, bei_predicting_2021, bracher_endemic-epidemic_2022, briz-redon_impact_2021, briz-redon_comparison_2022, briz-redon_association_2022, carroll_community_2021, chen_spatio-temporal_2023, chirombo_prevalence_2024, epstein_mapping_2023, freitas_spatio-temporal_2021, griffith_spatial-temporal_2021, ibeji_bayesian_2022, iddrisu_spatio-temporal_2022, jalilian_hierarchical_2021, jeong_investigating_2023, li_spatio-temporal_2023, liu_updated_2021, liu_influence_2023, liu_role_2025, lopez-lacort_bayesian_2024, nguyen_impact_2023, nguyen_modelling_2023, ngwira_spatial_2021, orozco-acosta_scalable_2023, pasanen_spatio-temporal_2024, petrof_using_2024, pirani_effects_2024, satorra_bayesian_2024, sun_spatio-temporal_2021, tam_bayesian_2024, yin_bayesian_2024, yip_spatio-temporal_2022, zhang_high-resolution_2024, zhang_bayesian_2025}\\ 
  Ecology          &   20       &  \cite{choi_short-term_2021, choi2025spatiotemporal, dambrine_characterising_2021, fioravanti_spatio-temporal_2021, freshwater_integrated_2021, garcia_bayesian_2023, gomez_integrating_2021, gruss_integrating_2023, huang_driving_2021, lindmark_evaluating_2023, lozano_spatio-temporal_2023, mcdonald_integrating_2023, neupane_novel_2022, olmos_estimating_2023, sando_multivariate_2024, staeudle_accounting_2024, tepe_history_2024, valente_pre-harvest_2021, zhang_inferring_2022, zhao_dynamic_2022}\\ 
  Public Health    &   10       &  \cite{beloconi_spatio-temporal_2021, bofa_optimizing_2024, bofa_bayesian_2024, he_identifying_2021, lome-hurtado_patterns_2021, ma_bayesian_2023, martenies_spatiotemporal_2021, nnanatu_evaluating_2021, saez_spatial_2022, wang_spatio-temporal_2025}\\ 
  Economics        &   6        &  \cite{baldoni_agricultural_2021, costantino_spatial_2023, elhorst_spatial_2022, mattera_forecasting_2025, pietak_effect_2024, yan_how_2025}\\ 
  Criminology      &   6        &  \cite{briz-redon_bayesian_2024, clark_class_2021, gracia_chronic_2021, mahfoud_forecasting_2021, vicente_multivariate_2023, vomfell_no_2023}\\
  Others\footnotemark &  5      &  \cite{adams_learning_2023, chen_bayesian_2024, maia_migration_2021, racek_capturing_2025, zaouche_bayesian_2023}\\ \hline
\end{tabularx}
\caption{Application area distribution of the included sources.}\label{tab:appl.areas}
\end{table}
\footnotetext{Shape modeling (1), Electronic vehicle charging (1), Migration and Electoral Participation (1), Armed Conflicts (1), Traffic (1)}

\subsubsection{Purpose of model fitting.} 
Across all fields, we identify four recurring categories of motivation: (i) description and spatio-temporal mapping, (ii) prediction and early warning, (iii) analysis of external determinants, and (iv) methodological development and data integration.

In epidemiology and public health research, the primary focus is on recording disease dynamics, monitoring epidemiological patterns, and analyzing health inequalities and exposures (detailed in Appendices \ref{subsub:Epidemiology} and \ref{subsub:PublicHealth}). In ecology, on the other hand, the focus is on modeling environmental and species changes, often with reference to nature conservation and policy issues (detailed in Appendix \ref{subsub:Ecology}). Economics primarily uses spatio-temporal approaches to study growth and productivity differences, predict economic indicators, and analyze regional interdependencies (detailed in Appendix \ref{subsub:Economics}). In the field of criminology, they are mainly used to overcome data uncertainty, predict criminal events, and analyze social determinants of violence (detailed in Appendix \ref{subsub:Crime}). 

\subsubsection{Model Structures.}
We present the model structures we found in the reviewed papers in the following.

\textit{Model Architecture.} As depicted in our full model structure overview in table \ref{tab:classification-models}, our literature search shows that hierarchical models are used significantly more often than flat models. We looked at 60 hierarchical models and 26 flat models. In the fields of epidemiology, ecology, and public health, hierarchical models are used more frequently than flat models (see Tables \ref{tab:models-epidem}, \ref{tab:models-ecology}, and \ref{tab:models-publichealth}). A particularly extreme example is the field of criminology, where we found only hierarchical models in the reviewed papers (see Table \ref{tab:models-crime}). The other extreme is found in economics, where no hierarchical models are used (see Table \ref{tab:models-economics}). 

\textit{Model Characteristics and Specification - Additive Spatio-Temporal Structure.} As we show in Table \ref{tab:classification-models}, the most common model characteristics are the additive spatio-temporal structure (detailed in Section \ref{subsub:L2-3} \ref{subsec:ST-CoV}). We find 64 models that use such an additive spatio-temporal component. Additive structures are used in at least half of the models in all categories. An extreme case here is the field of public health, where all reviewed models use additive spatio-temporal components (see Table \ref{tab:models-publichealth}).  

The most frequently found specification of an additive spatial component is the BYM specification (see Table \ref{tab:spatialEffect-structures}). We find this in 18 cases. This specification occurs predominantly in the field of epidemiology (13 cases). We find no BYM specification in the fields of ecology and economics. Furthermore, we find an FE specification of the additive spatial component in a total of 12 cases (see Table \ref{tab:spatialEffect-structures}). While this occurs less frequently in epidemiology (one case), the FE specification is the preferred modeling approach for the additive spatial component in economics. Here, it is chosen in five of the six models. 

For specifying the additive temporal component, the RW specification (17 cases) and the i.i.d. normal distribution specification (14 cases) are most frequently chosen (see Table \ref{tab:temporalEffect-structures}). In six cases, both occur together, with the temporal component being divided additively into a structured and an unstructured random effect. It is also striking that the field of epidemiology contains the largest proportion of such model specifications. Here we find 13 of the 17 RW specifications and eight of the 14 i.i.d. normal distribution specifications, with four containing the aforementioned combination. In eight models, an FE specification is chosen for the temporal component (see Table \ref{tab:temporalEffect-structures}). These occur primarily in the fields of economics and ecology, whereas such specifications are not found in the field of epidemiology. 

The spatio-temporal additive component is specified in 18 out of 34 cases with a separable covariance function. We find 12 of them in the application domain of epidemiology, three in public health research, two in the area of ecology, and one in the field of criminology. In total, we find nine models with the specification of the spatio-temporal additive component as a temporal autoregressive structure with spatial correlation innovation. We found three of this type of specification in epidemiology and public health, respectively. Two models of the area ecology and one model classified into the application domain \textit{others} use this specification of the additive spatio-temporal component. We observe that in the economic application domain, no spatio-temporal additive structure is employed.

\textit{Model Characteristics and Specification - Lag Structure.} Spatial, temporal, and spatio-temporal lag structures (detailed in Section \ref{subsub:L2-3} \ref{subsec:lag-structures}) are used in 22 cases (see Table \ref{tab:classification-models}). Lag structures can be found in all areas of application, except for the field of criminology. Another extreme case is the field of economics, where lag structures are used in all reviewed models (see Table \ref{tab:models-economics}). 

Regarding the specification of lag structures, we find spatial lag structures in 16 cases and temporal lag structures in 15 cases in the examined models. A spatio-temporal lag structure is used in seven cases. A detailed specification can be found in Table \ref{tab:Lag-structures}. While the lag structures used in epidemiology are primarily limited to one specification, e.g., only temporal lags \cite{zhang_bayesian_2025} or only spatial lags with regard to the covariates \cite{barcelo_spatiotemporal_2024}, all models in economics employ combinations of different lag structures. For example, \cite{baldoni_agricultural_2021} uses spatial lags with respect to the covariates in addition to temporal lags in the target variable and in the error term. We also note that lag structures most commonly occur in combination with additive spatio-temporal structures. We find such a combination in 16 out of 22 cases. 

\textit{Model Characteristics and Specification - Point Processes and Others.} In the publications reviewed, spatio-temporal point processes are used in the fields of epidemiology, ecology, and criminology. We found a total of four applications. We only find the application of a stochastic compartment model in the field of epidemiology. A geographically and temporally weighted regression (GTWR) approach is used in both epidemiology and the modeling of users' overstaying behavior at electric vehicle charging stations in \cite{chen_bayesian_2024}, classified under \textit{others}. Domain-adapted state space models and the two spatio-temporal extreme value models are used in the field of ecology.   

\subsubsection{Limitations.}
Across spatio-temporal studies, we can group the reported challenges and limitations into three categories: (i) data quality and availability, (ii) model assumptions and structure, and (iii) methodological constraints.

Data quality and availability are frequently cited as major limitations. Many studies report issues such as underreporting \cite{amaral_spatio-temporal_2023, fioravanti_spatio-temporal_2021, freitas_spatio-temporal_2021, garcia_bayesian_2023, huang_driving_2021, neupane_novel_2022}, misreporting \cite{barcelo_spatiotemporal_2024, iddrisu_spatio-temporal_2022}, missing data \cite{freshwater_integrated_2021, lindmark_evaluating_2023, liu_updated_2021, ngwira_spatial_2021, pasanen_spatio-temporal_2024, staeudle_accounting_2024}, and uncertainties in measurements \cite{freshwater_integrated_2021, olmos_estimating_2023}. Studies also mention general data availability issues \cite{bofa_bayesian_2024, he_identifying_2021, nguyen_modelling_2023, pietak_effect_2024, yip_spatio-temporal_2022}. Particularly, studies report challenges such as reliance on sparse or simulated data \cite{choi_short-term_2021, dambrine_characterising_2021, gomez_integrating_2021, zhao_dynamic_2022}, on limited historical records \cite{martenies_spatiotemporal_2021}, or on self-reported data \cite{nnanatu_evaluating_2021}. Some studies face biases in secondary or survey data \cite{iddrisu_spatio-temporal_2022, petrof_using_2024, zhang_high-resolution_2024}, or work with aggregated data at coarse spatial resolution \cite{freitas_spatio-temporal_2021, lome-hurtado_patterns_2021}. Different data limitations are prominent in the field of criminology, for example, the reliance on official records that may not reflect the full scope of criminal activity \cite{gracia_chronic_2021}. 

Model assumptions and structure represent another significant limitation. Many models are based on simplified assumptions that may not adequately reflect the true underlying mechanisms \cite{amaral_spatio-temporal_2023, bei_predicting_2021, briz-redon_impact_2021, briz-redon_association_2022, liu_role_2025}. Such assumptions can compromise both predictive performance and causal inference, mainly when key factors such as unobserved confounders, edge effects, or spatio-temporal interactions are not adequately accounted for \cite{briz-redon_comparison_2022, zhang_bayesian_2025}. 
Many models rely on assumptions that spatial or temporal structures are independent or separable \cite{sando_multivariate_2024, saez_spatial_2022}. Other studies report simplified assumptions about the underlying dynamics. For example, the assumption of linear dynamics in land development \cite{tepe_history_2024} or the assumption of purely local spillovers \cite{briz-redon_bayesian_2024, clark_class_2021}. Particularly, studies in economics rely on predefined spatial weight matrices \cite{baldoni_agricultural_2021, pietak_effect_2024}, geographical distance-based structures \cite{mattera_forecasting_2025}, or assume temporal stationarity, all of which may fail to capture dynamic spatial dependencies. Several studies also rely on assumptions regarding data structures, such as the use of logistic regression for temporal uncertainty, the reliance on smoothing-based assumptions \cite{dambrine_characterising_2021}, or the use of proxies \cite{martenies_spatiotemporal_2021, vomfell_no_2023}. Simplifications may reduce interpretability. Examples include focusing solely on morphological rather than functional or network-based spatial linkages \cite{yan_how_2025}, omitting boundary regions \cite{elhorst_spatial_2022}, or assuming that global factors dominate over local cluster-specific dynamics \cite{mattera_forecasting_2025}. Furthermore, some models are limited by their theoretical design and are not suitable for specific tasks, such as forecasting \cite{zhang_inferring_2022} or scaling across contexts \cite{mcdonald_integrating_2023, tepe_history_2024}. 

Methodological constraints restrict study design and implementation. Some models require pre-specification of the number of clusters \cite{ma_bayesian_2023} or rely heavily on informative Bayesian priors \cite{bofa_optimizing_2024, bofa_bayesian_2024, liu_updated_2021}, which may influence outcomes. Moreover, methodological challenges also arise in part from the integration of multiple data sources \cite{gruss_integrating_2023, staeudle_accounting_2024}. Additional challenges include computational limitations in handling complex model structures \cite{clark_class_2021, elhorst_spatial_2022, mahfoud_forecasting_2021, orozco-acosta_scalable_2023, vicente_multivariate_2023}, as well as from trade-offs between mesh resolution and computational cost \cite{staeudle_accounting_2024}. Context-specific factors may also influence model performance \cite{neupane_novel_2022}.

%%%%%%%%%%%%%%%%%%%%%%%%%%%%%%%%%%%%%%%%%%%%%% Discussion
\section{Discussion}\label{sec:discussion}
We structure the discussion of our work as follows: First, we will discuss the general results and findings of our literature review in Section \ref{subsec:discussion-Interpretation}. Then, we address in Section \ref{subsec:discussion-RQ} the research questions that form the basis of our review (see Section \ref{sec:introduction}). We conclude with the limitations of our work in Section \ref{subsec:discussion-Limitations}.

\subsection{General Results and Findings}\label{subsec:discussion-Interpretation}
\begin{figure}[H]
    \centering
    \includegraphics[width=\linewidth]{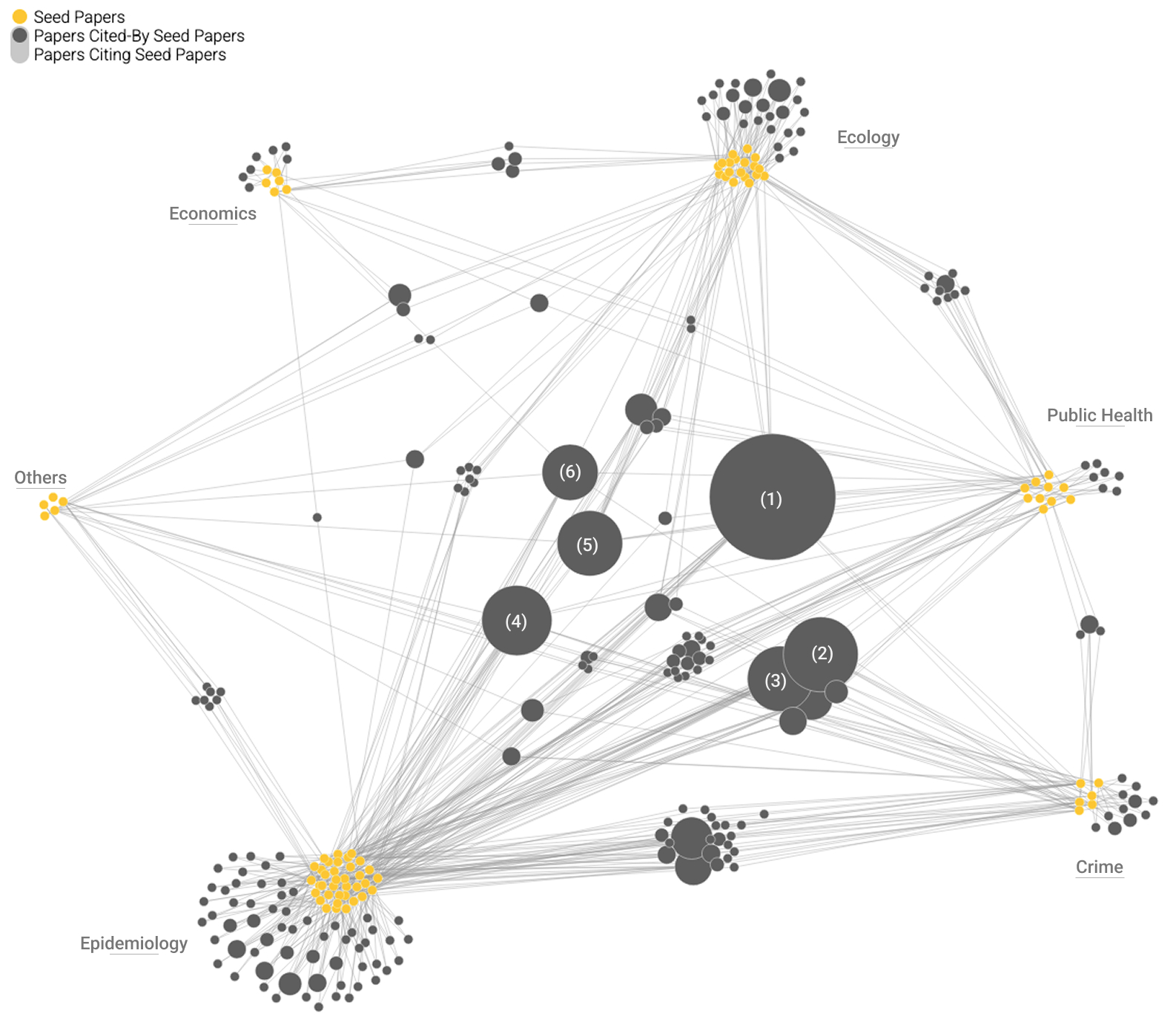}
    \caption{Graph of first-level citations that are cited more than once in the reviewed articles. The articles we reviewed are marked yellow and are placed according to the labeled application domains. The larger the dot, the more frequently the corresponding work is cited. The citation graph was constructed using Citation Gecko \url{https://citationgecko.azurewebsites.net/}. The discussion can be found in Section \ref{subsec:discussion-Interpretation}.}
    \label{fig:citationGraph}
\end{figure}

Our contribution is twofold: On the one hand, we propose a simple classification scheme (see Section \ref{subsec:modelling_techniques}) for spatio-temporal model structures. Our classification scheme provides a simplified overview of the most commonly used model structures for modeling spatio-temporal dependencies, making it easier for researchers to classify existing models. 
On the other hand, we examine the application of these structures in various areas (see Section \ref{subsec:application-domains}). 
We demonstrate that spatio-temporal data are widely used across various domains, and many real-world phenomena exhibit both spatial interactions and temporal dynamics. We find similar model structures in many applications (see Section \ref{subsec:application-domains}). However, our results also show that individual research areas could include more relevant publications from distinct application areas (see also the discussion of (RQ1) in Section \ref{subsec:discussion-RQ}). We created a citation graph, shown in Figure \ref{fig:citationGraph}, for the first-level citations that are cited more than once in the reviewed articles. We see that the papers cited by several application domains are methodological and theoretical standard works (see Table \ref{tab:CitedPapers}). This citation connection implies a rough consensus in methodological and theoretical guidelines across several application fields. However, we also note that the field of economics seems to be less strongly connected to the others, since it only appears once in citing the standard works in Table \ref{tab:CitedPapers}. Although ecology is the second largest application area in our overview in terms of the number of articles included (see Appendix \ref{subsub:Ecology}), only three of the six standard works are cited (see Table \ref{tab:CitedPapers}), which is comparatively few, especially in contrast to smaller areas such as criminology. Furthermore, there is also only one direct citation between the reviewed papers across different application domains. One paper \cite{saez_spatial_2022}, which we labeled as an application in the domain of public health, cites the paper \cite{fioravanti_spatio-temporal_2021} from the domain of ecology.
\begin{table}[h]
\footnotesize
\begin{tabularx}{\linewidth}{p{1.9cm}X}
\hline
 Reference                           & Cited by \\ \hline
\cite{rue2009approximate}  (marked as (1) in Figure \ref{fig:citationGraph})   &  Epidemiology: \cite{amaral_spatio-temporal_2023, barcelo_spatiotemporal_2024, briz-redon_impact_2021, briz-redon_comparison_2022, briz-redon_association_2022, carroll_community_2021, iddrisu_spatio-temporal_2022, jalilian_hierarchical_2021, li_spatio-temporal_2023, orozco-acosta_scalable_2023, petrof_using_2024, pirani_effects_2024, satorra_bayesian_2024, sun_spatio-temporal_2021, yin_bayesian_2024, yip_spatio-temporal_2022}, Ecology: \cite{dambrine_characterising_2021, fioravanti_spatio-temporal_2021, lindmark_evaluating_2023, staeudle_accounting_2024, valente_pre-harvest_2021}, Criminology: \cite{clark_class_2021, mahfoud_forecasting_2021, vicente_multivariate_2023}, Public Health: \cite{beloconi_spatio-temporal_2021, saez_spatial_2022, wang_spatio-temporal_2025}           \\
\cite{besag1991bayesian} (marked as (2) in Figure \ref{fig:citationGraph})    &  Epidemiology: \cite{ briz-redon_impact_2021, briz-redon_comparison_2022, briz-redon_association_2022, chen_spatio-temporal_2023, ibeji_bayesian_2022, iddrisu_spatio-temporal_2022, jalilian_hierarchical_2021, jeong_investigating_2023, ngwira_spatial_2021, satorra_bayesian_2024, yip_spatio-temporal_2022}, Criminology: \cite{briz-redon_bayesian_2024, clark_class_2021, gracia_chronic_2021, mahfoud_forecasting_2021}, Public Health: \cite{ma_bayesian_2023}           \\
\cite{knorr2000bayesian} (marked as (3) in Figure \ref{fig:citationGraph})   & Epidemiology: \cite{amaral_spatio-temporal_2023, briz-redon_impact_2021, carroll_community_2021, chen_spatio-temporal_2023, ibeji_bayesian_2022, lopez-lacort_bayesian_2024, ngwira_spatial_2021, orozco-acosta_scalable_2023, petrof_using_2024, satorra_bayesian_2024, yip_spatio-temporal_2022}, Criminology: \cite{gracia_chronic_2021, vicente_multivariate_2023}, Public Health: \cite{bofa_optimizing_2024}              \\ 
\cite{spiegelhalter2002bayesian} (marked as (4) in Figure \ref{fig:citationGraph})  &  Epidemiology: \cite{briz-redon_impact_2021, briz-redon_comparison_2022, briz-redon_association_2022, jeong_investigating_2023, ngwira_spatial_2021, orozco-acosta_scalable_2023, petrof_using_2024, pirani_effects_2024, yip_spatio-temporal_2022}, Criminology: \cite{briz-redon_bayesian_2024, mahfoud_forecasting_2021, vicente_multivariate_2023}, Public Health: \cite{beloconi_spatio-temporal_2021, nnanatu_evaluating_2021}, Others: \cite{zaouche_bayesian_2023}            \\
\cite{lindgren2011explicit} (marked as (5) in Figure \ref{fig:citationGraph})  &  Ecology: \cite{dambrine_characterising_2021, fioravanti_spatio-temporal_2021, gruss_integrating_2023, lindgren2011explicit, mcdonald_integrating_2023, olmos_estimating_2023, staeudle_accounting_2024, valente_pre-harvest_2021}, Public Health: \cite{beloconi_spatio-temporal_2021, saez_spatial_2022, wang_spatio-temporal_2025}, Epidemiology: \cite{li_spatio-temporal_2023, yin_bayesian_2024}, Others: \cite{zaouche_bayesian_2023}        \\
\cite{lindgren2015bayesian} (marked as (6) in Figure \ref{fig:citationGraph})  &  Epidemiology: \cite{briz-redon_impact_2021, briz-redon_comparison_2022, briz-redon_association_2022, jalilian_hierarchical_2021, pirani_effects_2024, satorra_bayesian_2024, zhang_high-resolution_2024}, Public Health: \cite{saez_spatial_2022}, Ecology: \cite{staeudle_accounting_2024}, Economics: \cite{baldoni_agricultural_2021}, Criminology: \cite{vicente_multivariate_2023}, Others: \cite{zaouche_bayesian_2023}           \\\hline
\end{tabularx}
\caption{The most cited papers in the reviewed articles. The dots illustrating paper (1) to (6) and we show their connections in Figure \ref{fig:citationGraph}.}\label{tab:CitedPapers}
\end{table}   
We assume that researchers can effectively reuse methodological approaches. Therefore, we should advocate for greater openness, as this is beneficial for science in general \cite{spellman2017open}. This view also leads us to criticize that no software code is publicly available for 47 of the reviewed publications. For one publication, only partial code is available, and another paper refers the reader to a publication scheduled for a later date. Code is available for only 34 publications. However, for five papers, it is only available upon request.

\subsection{Discussion of the Research Questions}\label{subsec:discussion-RQ}
\textit{(RQ1) What are the most commonly used spatio-temporal structures in statistical research, and how frequently are they applied across domains?}
In Section \ref{subsec:application-domains}, we see that the few model structures available for modeling spatio-temporal dependencies are used in all application areas. However, we also see that specific model characteristics and specifications are preferred in different application areas. On the one hand, this is because the model structures heavily depend on the availability of data (e.g., the aggregation of spatial and temporal units). 
On the other hand, we can observe that researchers often prioritize approaches that align with established methods within their respective disciplines (see also Section \ref{subsec:discussion-Interpretation}).
To see this, we categorized the models according to their different fields of application (Section \ref{subsec:application-domains}). As we have already described in Section \ref{subsec:application-domains}, the use of the flat and the hierarchical model architecture is unevenly distributed across the individual fields of application. We also found that hierarchical models are used significantly more often than flat models overall. One possible explanation for the more frequent use of hierarchical models is that many fields of application, especially epidemiology, public health, and criminology, often work with data that have a natural nesting or multiple levels of analysis. Hierarchical models are particularly well suited to represent such structures adequately \cite{cressie2011statistics}. In contrast, economics often uses models that focus more on aggregate variables or clearly defined economic units, with the emphasis frequently on causal identification and structural models. These requirements can often be met using flat models, which could explain the limited use of hierarchical approaches in this field. 

What we also noticed when reviewing the papers is that Bayesian statistics are used in 56 of the 83 publications. Except for economics, Bayesian statistics are used more frequently than the classical frequentist approach in all fields of application. We assume that the Bayesian approach will be a promising future field of research for spatio-temporal statistical models (see e.g., \cite{wikle2023statistical}).

\textit{(RQ2) In which domains are spatio-temporal models currently applied, and for what types of problems?}
We have demonstrated in our review that spatio-temporal statistical models are used for a wide range of tasks (see Section \ref{subsec:application-domains}). The widespread use cases suggest that spatio-temporal statistical models offer a powerful method for analyzing spatio-temporal data. In Appendix \ref{apx:application-domains} we describe five main application domains of spatio-temporal statistical models: Epidemiology, ecology, public health, economics, and criminology. These areas differ significantly and also pursue distinct goals (see Section \ref{subsec:application-domains}). Although spatio-temporal models are mainly established in the application areas considered in Section \ref{subsec:application-domains}, some application tasks in other areas involve spatial and temporal dependencies. Examples include migration \cite{maia_migration_2021}, shape modeling \cite{adams_learning_2023}, and traffic \cite{zaouche_bayesian_2023}. However, our literature review also reveals areas where spatio-temporal statistical models have not yet been applied. One possible explanation for this is that data availability in some areas is insufficient. For example, in media research for analyzing media bias \cite{Spinde2021f, SPINDE2023100264} or social media structures \cite{koutsaki2023spatiotemporal}. Future research should focus on broadening the application of spatio-temporal models through improved data accessibility and interdisciplinary exchange.

\textit{(RQ3) What are the key challenges and limitations in the application of spatio-temporal models across different domains?}
General conclusions: We can identify a tension between model complexity, data availability, and generalizability in all application domains. We suspect that data quality and availability are key barriers to the application of spatio-temporal models across all disciplines. Simplified model assumptions and the difficulty of capturing complex, non-linear, or dynamic processes remain common challenges. Methodological constraints also persist, including limited computational power and difficulties in integrating large, heterogeneous datasets. Common challenges and limitations are summarized in Section \ref{subsec:application-domains}.

Discipline-specific conclusions: Epidemiology and public health require models that can better capture the dynamics of intervention strategies and population mobility. Ecological models must account for more complex, non-linear ecological interactions and extreme events. Economics and criminology require more sophisticated models to capture network interactions and social dynamics. To see this, we identified the discipline-specific challenges in Appendix \ref{subsub:Epidemiology} to \ref{subsub:Crime}. 

The discussed challenges highlight the need to further enhance both the quality of the data and the model structures to capture the complexity of real phenomena better and increase the interpretability and predictive power of the models. Future research should focus more on developing more robust models and integrating innovative methods, such as big data technologies, to overcome these limitations.

\subsection{Limitations}\label{subsec:discussion-Limitations}
Since the structure of a spatio-temporal statistical model is crucial for interpreting the inference results, our literature review focuses exclusively on model structures that model spatial interaction and temporal dynamics. As we concentrate on this aspect, one major limitation of our literature review is that we do not consider areas such as spatio-temporal descriptive statistics, parameter estimation, or validation criteria for spatio-temporal models. Furthermore, we do not analyze criteria for model structure selection.

%%%%%%%%%%%%%%%%%%%%%%%%%%%%%%%%%%%%%%%%%%%%%% Conclusion
\section{Conlusion}\label{sec:conclusion}
We conducted a systematic literature review according to the PRISMA scheme and included 83 publications in our evaluation. To classify the models described in the included papers, we propose a scheme for categorizing spatio-temporal statistical model structures. We show that (1) although only a few frequently used model structures are employed in a wide range of application areas, (2) preferences can be identified in the individual application areas. Our review identifies shared theoretical foundations and divergent practices, providing a basis for tailored approaches that enhance informed decision-making across various fields. It also motivates interdisciplinary collaboration, which helps identify the most effective models for specific problems.

\newpage
\section*{Appendix}
\section{Tables and Figures}

\begin{table}[h]
\footnotesize
\begin{listliketab}
  \begin{tabularx}{\linewidth}{@{}>{\bfseries}l @{\hspace{.5em}} XR}
%\begin{tabular}{|l|l|}
\hline
Database & Search String \\ \hline
  Scopus       &          TITLE-ABS-KEY ( "spatio-temporal model*" OR "spatio-temporal process*" OR "spatial-temporal model*" OR "statistical analysis of spatio-temporal data" OR "statistical analysis of space time data" OR "spacetime statistical model" OR "geo-temporal model" OR "geostatistical space time model*" OR "spatio-temporal point process" OR "spatio-temporal Gaussian process model" OR "Bayesian spatio-temporal modeling" OR "dynamic spatio-temporal model*" OR "spatial panel data model*" OR "spatio-temporal regression" OR "spatio-temporal mixed model*" ) AND PUBYEAR > 2020 AND ( LIMIT-TO ( LANGUAGE , "English" ) ) AND ( LIMIT-TO ( EXACTKEYWORD , "Spatio-temporal Models" ) OR LIMIT-TO ( EXACTKEYWORD , "Spatiotemporal Analysis" ) ) AND ( LIMIT-TO ( OA , "all" ) )     \\ \hline
  Web of Science       &      TS=("spatio-temporal model*" OR "spatio-temporal process*" OR "spatial-temporal model*" OR "statistical analysis of spatio-temporal data" OR "statistical analysis of space time data" OR "spacetime statistical model" OR "geo-temporal model" OR "geostatistical space time model*" OR "spatio-temporal point process" OR "spatio-temporal Gaussian process model" OR "Bayesian spatio-temporal modeling" OR "dynamic spatio-temporal model*" OR "spatial panel data model*" OR "spatio-temporal regression" OR "spatio-temporal mixed model*")
AND PY=(2021-2025)
AND LA=(English)         \\ \hline
%\end{tabular}
\end{tabularx}
\end{listliketab}
\caption{Databases and corresponding search string.}\label{tab:searchString}
\end{table}

\begin{landscape}

\begin{table}[]
\footnotesize
\begin{tabular}{l|cccccccc}
  & \multicolumn{1}{c|}{\textbf{Application Domain}} & \multicolumn{1}{c|}{\textbf{Statistic}} & \multicolumn{2}{c|}{\textbf{Model Architecture}} & \multicolumn{4}{c}{\textbf{Model Characteristics}} \\
  & \multicolumn{1}{c|}{}  & \multicolumn{1}{c|}{}          & Flat     & \multicolumn{1}{c|}{Hierarchical}    & Additive Component(s)  & Lag Structure & Intensity Function & Others \\ \hline
 
 \cite{baldoni_agricultural_2021}       & Economics     & B & x &  & x  & x &   & \\
 \cite{bei_predicting_2021}             & Epidemiology  & F &  & x &    &   & x & \\
 \cite{beloconi_spatio-temporal_2021}   & Pub. Health   & B & x &  & x  & x &   & \\
 \cite{beloconi_spatio-temporal_2021}   & Pub. Health   & B & x &  & x  &   &   & \\
 \cite{briz-redon_impact_2021}          & Epidemiology  & B &  & x & x  &   &   & \\
 \cite{carroll_community_2021}          & Epidemiology  & B &  & x & x  &   &   & \\
 \cite{choi_short-term_2021}            & Ecology       & B &  & x &    &   &   & x \\
 \cite{clark_class_2021}                & Criminology   & B &  & x &    &   & x &  \\
 \cite{dambrine_characterising_2021}    & Ecology       & B &  & x & x  &   &   & \\
 \cite{fioravanti_spatio-temporal_2021} & Ecology       & B & x &  & x  &   &   &\\
 \cite{freitas_spatio-temporal_2021}    & Epidemiology  & B &  & x & x  &   &   & \\
 \cite{freshwater_integrated_2021}      & Ecology       & F &  & x & x  &   &   & \\
 \cite{gomez_integrating_2021}          & Ecology       & B &  & x &    &   &   & x\\
 \cite{gracia_chronic_2021}             & Criminology   & B &  & x & x  & x &   &  \\
 \cite{griffith_spatial-temporal_2021}  & Epidemiology  & F &  & x & x  &   &   &\\
 \cite{griffith_spatial-temporal_2021}  & Epidemiology  & F &  & x & x  &   &   & \\
 \cite{he_identifying_2021}             & Pub. Health   & F & x &  & x  & x &   & \\
 \cite{huang_driving_2021}              & Ecology       & B & x &  & x  & x &   &\\
 \cite{jalilian_hierarchical_2021}      & Epidemiology  & B &  & x & x  &   &   & \\
 \cite{liu_updated_2021}                & Epidemiology  & F &  & x & x  &   &   & \\
 \cite{lome-hurtado_patterns_2021}      & Pub. Health   & B &  & x & x  &   &   & \\
 \cite{mahfoud_forecasting_2021}        & Criminology   & B &  & x & x  &   &   & \\
 \cite{maia_migration_2021}             & Others        & F & x &  & x  & x &   & \\
 \cite{martenies_spatiotemporal_2021}   & Pub. Health   & F & x &  & x  &   &   & \\
 \cite{ngwira_spatial_2021}             & Epidemiology  & B &  & x & x  &   &   & \\
 \cite{nnanatu_evaluating_2021}         & Pub. Health   & B &  & x & x  &   &   & \\
 \cite{sun_spatio-temporal_2021}        & Epidemiology  & B &  & x & x  &   &   & \\
 \cite{valente_pre-harvest_2021}        & Ecology       & B &  & x &    &   & x &\\ 
 \cite{bracher_endemic-epidemic_2022}   & Epidemiology  & F &  & x &    & x &   &\\
 \cite{briz-redon_association_2022}     & Epidemiology  & B & x &  & x  & x &   & \\ 
 \cite{briz-redon_comparison_2022}      & Epidemiology  & B &  & x & x  &   &   &\\
 \cite{elhorst_spatial_2022}            & Economics     & F & x &  & x  & x &   & \\
 \cite{ibeji_bayesian_2022}             & Epidemiology  & B &  & x & x  &   &   & \\
 \cite{iddrisu_spatio-temporal_2022}    & Epidemiology  & B &  & x & x  &   &   &\\
 \cite{neupane_novel_2022}              & Ecology       & B & x &  & x  &   &   & \\
 \cite{saez_spatial_2022}               & Pub. Health   & B &  & x & x  &   &   & \\ 
 \cite{yip_spatio-temporal_2022}        & Epidemiology  & B &  & x & x  & x &   &\\
 \cite{zhang_inferring_2022}            & Ecology       & F & x &  &    & x &   &\\ 
 \cite{zhao_dynamic_2022}               & Ecology       & F & x &  & x  & x &   & \\
 \cite{adams_learning_2023}             & Others        & F & x &  &    &   &   & x\\
 \cite{amaral_spatio-temporal_2023}     & Epidemiology  & B &  & x &    &   & x & \\
 \cite{chen_spatio-temporal_2023}       & Epidemiology  & B &  & x & x  &   &   &\\
 \hline
\end{tabular}
    \caption{Classification of the reviewed models according to our proposed classification scheme. If several models are described in one paper, they are listed below each other. The sources are listed in increasing order according to the publication year. B = Bayesian Approach , F = Frequentist Approach.}
    \label{tab:classification-models}
\end{table}

\begin{table}[]
\footnotesize
\begin{tabular}{l|cccccccc}
  & \multicolumn{1}{c|}{\textbf{Application Domain}} & \multicolumn{1}{c|}{\textbf{Statistic}} & \multicolumn{2}{c|}{\textbf{Model Architecture}} & \multicolumn{4}{c}{\textbf{Model Characteristics}} \\
  & \multicolumn{1}{c|}{}  & \multicolumn{1}{c|}{}          & Flat     & \multicolumn{1}{c|}{Hierarchical}    & Additive Component(s)  & Lag Structure & Intensity Function & Others \\ \hline

 \cite{costantino_spatial_2023}         & Economics     & F & x &  & x  & x &   & \\
 \cite{epstein_mapping_2023}            & Epidemiology  & F &  & x & x  &   &   & \\
 \cite{garcia_bayesian_2023}            & Ecology       & B &  & x &    &   &   & x \\
 \cite{gruss_integrating_2023}          & Ecology       & F &  & x & x  &   &   & \\
 \cite{jeong_investigating_2023}        & Epidemiology  & B &  & x & x  &   &   & \\
 \cite{li_spatio-temporal_2023}         & Epidemiology  & B &  & x & x  &   &   & \\
 \cite{lindmark_evaluating_2023}        & Ecology       & B &  & x & x  &   &   & \\
 \cite{lindmark_evaluating_2023}        & Ecology       & B &  & x & x  &   &   &\\
 \cite{liu_influence_2023}              & Epidemiology  & F & x &  &    &   &   & x \\
 \cite{lozano_spatio-temporal_2023}     & Ecology       & F & x &  & x  & x &   & \\
 \cite{ma_bayesian_2023}                & Pub. Health   & B &  & x & x  &   &   &  \\
 \cite{mcdonald_integrating_2023}       & Ecology       & F &  & x &    &   &   & x \\
 \cite{nguyen_impact_2023}              & Epidemiology  & F &  & x &    & x &   &\\
 \cite{nguyen_modelling_2023}           & Epidemiology  & B & x &  &    &   &   & x  \\
 \cite{olmos_estimating_2023}           & Ecology       & B &  & x & x  &   &   &\\
 \cite{orozco-acosta_scalable_2023}     & Epidemiology  & B &  & x & x  &   &   & \\
 \cite{vicente_multivariate_2023}       & Criminology   & B &  & x & x  &   &   & \\
 \cite{vomfell_no_2023}                 & Criminology   & F &  & x &    &   & x &\\
 \cite{zaouche_bayesian_2023}           & Others        & B &  & x & x  &   &   &\\
 \cite{barcelo_spatiotemporal_2024}     & Epidemiology  & B &  & x & x  & x &   &\\
 \cite{bofa_bayesian_2024}              & Pub. Health   & B & x &  & x  &   &   & \\
 \cite{bofa_optimizing_2024}            & Pub. Health   & B &  & x & x  &   &   &\\
 \cite{briz-redon_bayesian_2024}        & Criminology   & B &  & x & x  &   &   &\\
 \cite{chen_bayesian_2024}              & Others        & B & x &  &    &   &   & x\\
 \cite{chirombo_prevalence_2024}        & Epidemiology  & B &  & x & x  &   &   &\\
 \cite{lopez-lacort_bayesian_2024}      & Epidemiology  & B &  & x &    & x &   &\\
 \cite{pasanen_spatio-temporal_2024}    & Epidemiology  & B &  & x & x  &   &   &\\
 \cite{petrof_using_2024}               & Epidemiology  & B &  & x & x  &   &   &\\
 \cite{pietak_effect_2024}              & Economics     & F & x &  & x  & x &   &\\
 \cite{pirani_effects_2024}             & Epidemiology  & B &  & x & x  &   &   &\\
 \cite{sando_multivariate_2024}         & Ecology       & F & x &  &    &   &   & x\\
 \cite{satorra_bayesian_2024}           & Epidemiology  & B &  & x & x  &   &   & \\
 \cite{staeudle_accounting_2024}        & Ecology       & B &  & x & x  &   &   &\\
 \cite{tam_bayesian_2024}               & Epidemiology  & B &  & x & x  &   &   &\\
 \cite{tepe_history_2024}               & Ecology       & F & x &  &    & x &   & \\
 \cite{yin_bayesian_2024}               & Epidemiology  & B &  & x & x  &   &   &\\
 \cite{zhang_high-resolution_2024}      & Epidemiology  & B &  & x &x   &   &   &\\
 \cite{choi2025spatiotemporal}          & Ecology       & B &  & x &    &   &   & x\\
 \cite{liu_role_2025}                   & Epidemiology  & F & x &  &    &   &   & x\\ 
 \cite{mattera_forecasting_2025}        & Economics     & F & x &  &    & x &   &\\
 \cite{racek_capturing_2025}            & Others        & F &  & x & x  &   &   &\\
 \cite{wang_spatio-temporal_2025}       & Pub. Health   & B &  & x & x  &   &   &\\
 \cite{yan_how_2025}                    & Economics     & F & x &  & x  & x &   &\\
 \cite{zhang_bayesian_2025}             & Epidemiology  & B &  & x & x  & x &   &\\
 \hline
\end{tabular}
    \caption*{Table \ref{tab:classification-models} continued}
\end{table}

\end{landscape}

\section{Application Domains}\label{apx:application-domains}
We aim to link the findings on modeling strategies (discussed in Section \ref{subsec:modelling_techniques}) with the results specific to each application category. To this end, we describe each application area separately. For each, we discuss (i) the purpose of model fitting, (ii) spatio-temporal modeling strategies, and (iii) reported limitations. Section \ref{subsec:application-domains} summarizes the main findings and highlights similarities and differences across application areas. 

\subsection{Epidemiology.}\label{subsub:Epidemiology}
Epidemiology is the most represented field in our review, with a total of 36 publications classified under this category (see Table \ref{tab:appl.areas}).

\textit{Purpose of model fitting.}
A central aim in this field is to understand the spatio-temporal dynamics of diseases. Studies focus on identifying high-risk areas \cite{chirombo_prevalence_2024, iddrisu_spatio-temporal_2022}, explaining observed disease patterns \cite{amaral_spatio-temporal_2023, briz-redon_impact_2021, jeong_investigating_2023, liu_role_2025, nguyen_impact_2023, nguyen_modelling_2023, ngwira_spatial_2021, yin_bayesian_2024}, and supporting disease mapping and surveillance. The latter uses spatio-temporal models to monitor epidemiological patterns and guide resource allocation \cite{carroll_community_2021, epstein_mapping_2023, ibeji_bayesian_2022, liu_role_2025, lopez-lacort_bayesian_2024, zhang_bayesian_2025, zhang_high-resolution_2024}. 
Most studies in this group rely on hierarchical regression models, with Poisson regression being the most common \cite{briz-redon_impact_2021, carroll_community_2021, chirombo_prevalence_2024, ibeji_bayesian_2022, iddrisu_spatio-temporal_2022, ngwira_spatial_2021, yin_bayesian_2024, zhang_high-resolution_2024, zhang_bayesian_2025}. Other approaches include stochastic compartment models \cite{liu_role_2025, nguyen_modelling_2023} and point process models \cite{amaral_spatio-temporal_2023}. 

A second group of studies emphasizes prediction and early warning. Here, spatio-temporal models are applied to forecast new cases or impute missing data in health records \cite{bei_predicting_2021, bracher_endemic-epidemic_2022, griffith_spatial-temporal_2021, li_spatio-temporal_2023, liu_updated_2021, nguyen_modelling_2023, orozco-acosta_scalable_2023, yip_spatio-temporal_2022}. This predictive focus also extends to veterinary and environmental epidemiology, particularly in monitoring vector-borne diseases and zoonoses \cite{freitas_spatio-temporal_2021, liu_influence_2023, liu_role_2025, nguyen_modelling_2023, pirani_effects_2024, sun_spatio-temporal_2021}.
In this group, a variety of model classes are used. These include point process models \cite{bei_predicting_2021}, stochastic compartment models \cite{liu_role_2025, nguyen_modelling_2023}, and regression models \cite{bracher_endemic-epidemic_2022, freitas_spatio-temporal_2021, griffith_spatial-temporal_2021, li_spatio-temporal_2023, liu_updated_2021, liu_influence_2023, orozco-acosta_scalable_2023, pirani_effects_2024, sun_spatio-temporal_2021, yip_spatio-temporal_2022}. 

A third motivation is to explain the role of external factors that influence health. These studies integrate spatio-temporal structures with socioeconomic, demographic, climatic, or environmental covariates to investigate how such factors influence disease incidence and mortality \cite{barcelo_spatiotemporal_2024, briz-redon_impact_2021, briz-redon_association_2022, carroll_community_2021, chen_spatio-temporal_2023, jeong_investigating_2023, satorra_bayesian_2024, tam_bayesian_2024, yin_bayesian_2024}. This research highlights unequal disease burdens among socioeconomically disadvantaged populations \cite{barcelo_spatiotemporal_2024} and the effects of environmental change on the transmission of vector-borne diseases \cite{freitas_spatio-temporal_2021, pirani_effects_2024, sun_spatio-temporal_2021}. All models in this group are regression-based, predominantly Poisson regression models \cite{barcelo_spatiotemporal_2024, briz-redon_impact_2021, carroll_community_2021, chen_spatio-temporal_2023, freitas_spatio-temporal_2021, satorra_bayesian_2024, tam_bayesian_2024, yin_bayesian_2024}.

A final motivation lies in methodological refinement. Some studies propose or compare spatio-temporal models to address challenges such as sparse data in rural regions \cite{pasanen_spatio-temporal_2024} or inconsistencies across data sources \cite{petrof_using_2024}. Others examine the impact of varying model assumptions, such as neighborhood definitions or diffusion mechanisms, on inference and interpretation \cite{briz-redon_comparison_2022, li_spatio-temporal_2023}. This category exclusively includes regression models.

\textit{Model strategies.}
An overview of the modeling strategies used in epidemiology is provided in Table \ref{tab:models-epidem}. Hierarchical regression models dominate this application area, with Poisson regression being particularly prevalent. In total, 32 models employ a hierarchical structure, while only four do not. The latter include a segmented linear regression model \cite{briz-redon_association_2022}, a geographically and temporally weighted regression (GTWR) model \cite{liu_influence_2023}, and two stochastic compartment models \cite{liu_role_2025, nguyen_modelling_2023}.
Most models (28 in total) apply additive spatio-temporal structures (see Section \ref{subsec:ST-CoV}). Lag structures (see Section \ref{subsec:lag-structures}) are used in seven studies, with four of these combining lag structures with additive components \cite{barcelo_spatiotemporal_2024, briz-redon_association_2022, yip_spatio-temporal_2022, zhang_bayesian_2025}.
Two studies employ intensity functions in point process models to define spatio-temporal structure \cite{amaral_spatio-temporal_2023, bei_predicting_2021} (see Section \ref{subsec:PP}). Additionally, two studies use stochastic compartment models \cite{liu_role_2025, nguyen_modelling_2023}, and one uses a GTWR model \cite{liu_influence_2023}.
\begin{table}[h]
\footnotesize
\begin{tabularx}{\linewidth}{p{3cm}cX}
\hline
                        & Frequency  & References \\ \hline
  Hierarchical          &   32      &  \cite{amaral_spatio-temporal_2023, barcelo_spatiotemporal_2024, bei_predicting_2021, bracher_endemic-epidemic_2022, briz-redon_impact_2021, briz-redon_comparison_2022, carroll_community_2021, chen_spatio-temporal_2023, chirombo_prevalence_2024, epstein_mapping_2023, freitas_spatio-temporal_2021, griffith_spatial-temporal_2021, ibeji_bayesian_2022, iddrisu_spatio-temporal_2022, jalilian_hierarchical_2021, jeong_investigating_2023, li_spatio-temporal_2023, liu_updated_2021, lopez-lacort_bayesian_2024, nguyen_impact_2023, ngwira_spatial_2021, orozco-acosta_scalable_2023, pasanen_spatio-temporal_2024, petrof_using_2024, pirani_effects_2024, satorra_bayesian_2024, sun_spatio-temporal_2021, tam_bayesian_2024, yin_bayesian_2024, yip_spatio-temporal_2022, zhang_high-resolution_2024, zhang_bayesian_2025}\\ 
  Flat                  &   4       &  \cite{briz-redon_association_2022, liu_influence_2023, liu_role_2025, nguyen_modelling_2023}\\ \hline
  Additive Structures          &    28     &  \cite{barcelo_spatiotemporal_2024, briz-redon_impact_2021, briz-redon_comparison_2022, briz-redon_association_2022, carroll_community_2021, chen_spatio-temporal_2023, chirombo_prevalence_2024, epstein_mapping_2023, freitas_spatio-temporal_2021, griffith_spatial-temporal_2021, ibeji_bayesian_2022, iddrisu_spatio-temporal_2022, jalilian_hierarchical_2021, jeong_investigating_2023, li_spatio-temporal_2023, liu_updated_2021, ngwira_spatial_2021, orozco-acosta_scalable_2023, pasanen_spatio-temporal_2024, petrof_using_2024, pirani_effects_2024, satorra_bayesian_2024, sun_spatio-temporal_2021, tam_bayesian_2024, yin_bayesian_2024, yip_spatio-temporal_2022, zhang_high-resolution_2024, zhang_bayesian_2025}\\ 
  Lag Structures     &    7     &  \cite{barcelo_spatiotemporal_2024, bracher_endemic-epidemic_2022, briz-redon_association_2022, lopez-lacort_bayesian_2024, nguyen_impact_2023, yip_spatio-temporal_2022, zhang_bayesian_2025} \\ 
  Intensity Function &   2    &  \cite{amaral_spatio-temporal_2023, bei_predicting_2021}\\ 
  Others:      &        &  \\ 
  \hspace{0.05cm} Stoch. compartment model   &   2     &  \cite{liu_role_2025, nguyen_modelling_2023}\\
  \hspace{0.05cm} GTWR    &    1      &  \cite{liu_influence_2023}\\ \hline
\end{tabularx}
\caption{Models classified according to the proposed scheme in the application domain epidemiology}\label{tab:models-epidem}
\end{table} 

\textit{Limitations.}
Data quality and availability are frequently cited as significant limitations. Many studies report issues such as underreporting, misreporting, missing data, low temporal or spatial resolution, and biases in secondary or survey data \cite{amaral_spatio-temporal_2023, barcelo_spatiotemporal_2024, chirombo_prevalence_2024, epstein_mapping_2023, freitas_spatio-temporal_2021, iddrisu_spatio-temporal_2022, nguyen_modelling_2023, ngwira_spatial_2021, petrof_using_2024, yip_spatio-temporal_2022, zhang_high-resolution_2024}.

Model assumptions and structure represent another significant limitation. Many models are based on simplified assumptions that may not adequately reflect the true underlying mechanisms of disease dynamics \cite{amaral_spatio-temporal_2023, bei_predicting_2021, briz-redon_impact_2021, briz-redon_association_2022, liu_role_2025}. Such assumptions can compromise both predictive performance and causal inference.

Methodological constraints include limited temporal and spatial coverage, short study periods, aggregation over coarse spatial units, and computational challenges associated with high-dimensional hierarchical models \cite{griffith_spatial-temporal_2021, ibeji_bayesian_2022, liu_updated_2021, orozco-acosta_scalable_2023, pasanen_spatio-temporal_2024, yin_bayesian_2024}. Additional challenges stem from the inability to incorporate dynamic factors such as intervention timing, vector control, vaccination rollout, or population mobility. Furthermore, macroscopic or aggregated modeling approaches often restrict causal interpretation \cite{briz-redon_association_2022, satorra_bayesian_2024, sun_spatio-temporal_2021, tam_bayesian_2024}.

%%%%%%%%%%%%%%%%%%%%%%%%%%%%%%%%%%%%
\subsection{Ecology.}\label{subsub:Ecology}
The ecology application area is the second largest category in our review (see Table \ref{tab:appl.areas}).

\textit{Purpose of model fitting.}
One primary objective of using spatio-temporal models in ecological research is to monitor and predict environmental change. Several studies model environmental indicators such as near-surface temperature \cite{choi_short-term_2021}, Arctic sea ice \cite{zhang_inferring_2022}, and vegetation greenness \cite{neupane_novel_2022}. Others focus on extreme climate events \cite{garcia_bayesian_2023, sando_multivariate_2024} or provide high-resolution pollution and air quality maps \cite{fioravanti_spatio-temporal_2021}. For these purposes, both flat \cite{neupane_novel_2022, fioravanti_spatio-temporal_2021} and hierarchical \cite{garcia_bayesian_2023} regression models are employed. More complex structures are also used, including a domain-adapted state-space model \cite{choi_short-term_2021, choi2025spatiotemporal}, a spatio-temporal logistic auto-regressive model \cite{zhang_inferring_2022}, and a multivariate spatio-temporal extreme value and exposure model \cite{sando_multivariate_2024}.

A second central theme is the analysis of species dynamics and ecological interactions. These studies investigate the drivers of animal population dynamics and condition variability \cite{freshwater_integrated_2021, lindmark_evaluating_2023, mcdonald_integrating_2023, olmos_estimating_2023, staeudle_accounting_2024}, spawning areas \cite{dambrine_characterising_2021}, and responses to environmental pressures such as hunting and predation \cite{lozano_spatio-temporal_2023}. The spread of invasive species and pests under climate variability and extreme events is also addressed \cite{huang_driving_2021}. Most studies in this group use hierarchical regression models \cite{dambrine_characterising_2021, freshwater_integrated_2021, lindmark_evaluating_2023, olmos_estimating_2023, staeudle_accounting_2024}, along with one flat regression model \cite{lozano_spatio-temporal_2023} and one flat spatial panel data model \cite{huang_driving_2021}. An exception is the spatially explicit habitat-based assessment model in \cite{mcdonald_integrating_2023}, a domain-adapted state-space model used to integrate habitat features for a better understanding of sea scallop stock productivity.

A third key motivation is to assess the impact of human activities and policy interventions. Examples include studies on land-use change \cite{tepe_history_2024}, the effects of sugarcane burning regulations \cite{valente_pre-harvest_2021}, and the spillover effects of industrial policy on carbon emissions \cite{zhao_dynamic_2022}. Two of these studies use flat spatial panel data models incorporating spatial interaction and temporal dynamics \cite{tepe_history_2024, zhao_dynamic_2022}, while one employs a log-Gaussian Cox process \cite{valente_pre-harvest_2021}.

A further set of studies is motivated by the need for data integration and methodological refinement. These works aim to improve model accuracy and ecological inference by combining survey data with community science or qualitative sources \cite{choi2025spatiotemporal, gomez_integrating_2021, gruss_integrating_2023, mcdonald_integrating_2023}. All models in this group are hierarchical, though they vary in structure. They include a maximum entropy model \cite{gomez_integrating_2021}, a probabilistic spatio-temporal model, based on the stochastic advection–diffusion equation \cite{choi_short-term_2021}, a spatially explicit habitat-based assessment model \cite{mcdonald_integrating_2023}, and a Poisson regression model with additive spatio-temporal structure \cite{gruss_integrating_2023}.

\textit{Model strategies.}
An overview is provided in Table \ref{tab:models-ecology}. Of the 20 models reviewed, 12 follow a hierarchical architecture, while eight use a flat structure. Most models (11 in total) implement an additive spatio-temporal structure as described in Section \ref{subsec:ST-CoV}. Lag structures (see Section \ref{subsec:lag-structures}) appear in five models, three of which also incorporate additive components \cite{huang_driving_2021, lozano_spatio-temporal_2023, zhao_dynamic_2022}.
Three domain-adapted state-space models are used in the reviewed ecology studies \cite{choi_short-term_2021, choi2025spatiotemporal, mcdonald_integrating_2023}. Additionally, two spatio-temporal extreme value models appear in this domain \cite{garcia_bayesian_2023, sando_multivariate_2024}, reflecting the specific demands of ecological applications.
\begin{table}[h]
\footnotesize
\begin{tabularx}{\linewidth}{p{3.3cm}cX}
\hline
                        & Frequency  & References \\ \hline
  Hierarchical          &    12     &  \cite{choi_short-term_2021, choi2025spatiotemporal, dambrine_characterising_2021, freshwater_integrated_2021, garcia_bayesian_2023, gomez_integrating_2021, gruss_integrating_2023, lindmark_evaluating_2023, mcdonald_integrating_2023, olmos_estimating_2023, staeudle_accounting_2024, valente_pre-harvest_2021}\\ 
  Flat                  &     8     & \cite{fioravanti_spatio-temporal_2021, huang_driving_2021, lozano_spatio-temporal_2023, neupane_novel_2022, sando_multivariate_2024, tepe_history_2024,  zhang_inferring_2022, zhao_dynamic_2022}\\ \hline
  Additive Structures          &    11     & \cite{dambrine_characterising_2021, fioravanti_spatio-temporal_2021, freshwater_integrated_2021, gruss_integrating_2023, huang_driving_2021, lindmark_evaluating_2023, lozano_spatio-temporal_2023, neupane_novel_2022, olmos_estimating_2023, staeudle_accounting_2024, zhao_dynamic_2022} \\ 
  Lag Structures     &    5     &  \cite{huang_driving_2021, lozano_spatio-temporal_2023, tepe_history_2024, zhang_inferring_2022, zhao_dynamic_2022} \\ 
  Intensity Function &   1    &  \cite{valente_pre-harvest_2021} \\ 
  Others:      &        &  \\ 
  \hspace{0.05cm} State-space model   &   3     &  \cite{choi_short-term_2021, choi2025spatiotemporal, mcdonald_integrating_2023}\\
  \hspace{0.05cm} Extreme value model    &    2      &  \cite{garcia_bayesian_2023, sando_multivariate_2024}\\ \hline
\end{tabularx}
\caption{Models classified according to the proposed scheme in the application domain ecology}\label{tab:models-ecology}
\end{table}

\textit{Limitations.}
Data quality and availability frequently limit model performance. Many studies report challenges, including reliance on sparse or simulated data \cite{choi_short-term_2021, dambrine_characterising_2021, gomez_integrating_2021, zhao_dynamic_2022}, imperfect sampling \cite{freshwater_integrated_2021}, missing data \cite{staeudle_accounting_2024}, and uncertainties in measurements of catch, effort, or abundance \cite{freshwater_integrated_2021, olmos_estimating_2023}. Model accuracy and predictive performance are further constrained by limited observational data, particularly for extreme events or specific habitats \cite{fioravanti_spatio-temporal_2021, garcia_bayesian_2023, neupane_novel_2022}, or species with distinct generation cycles \cite{huang_driving_2021}.

Model assumptions and structural choices introduce additional challenges. Simplified assumptions, such as the independence of spatial exposures \cite{sando_multivariate_2024}, linear dynamics in land development \cite{tepe_history_2024}, or smoothing-based assumptions \cite{dambrine_characterising_2021}, may fail to reflect the complexity of ecological interactions. Furthermore, some models are limited by their theoretical design and are not suitable for specific tasks, such as forecasting \cite{zhang_inferring_2022} or scaling \cite{mcdonald_integrating_2023, tepe_history_2024}.

Methodological constraints arise in part from the integration of multiple data sources \cite{gruss_integrating_2023, staeudle_accounting_2024}, as well as from trade-offs between mesh resolution and computational cost \cite{staeudle_accounting_2024}. Model performance may also be influenced by context-specific factors, such as forest type \cite{neupane_novel_2022}. Additional limitations include the impact of unmeasured variables \cite{lindmark_evaluating_2023}, omitted effects such as seasonality \cite{gruss_integrating_2023}, and the lack of clearly defined treatment and control conditions in space and time \cite{valente_pre-harvest_2021}.

%%%%%%%%%%%%%%%%%%%%%%%%%%%%%%%%%%%%
\subsection{Public Health.}\label{subsub:PublicHealth}
We classify 10 papers as belonging to the field of public health (see Table \ref{tab:appl.areas}).

\textit{Purpose of model fitting.}
In public health research, spatio-temporal models are used to investigate environmental exposures and their health impacts. Several studies focus on developing models that capture spatio-temporal variability in pollutants, considering regulatory measures, socioeconomic factors, and computational efficiency \cite{beloconi_spatio-temporal_2021, he_identifying_2021, martenies_spatiotemporal_2021, saez_spatial_2022}. Most of these studies apply regression models \cite{beloconi_spatio-temporal_2021, martenies_spatiotemporal_2021, saez_spatial_2022}, while \cite{he_identifying_2021} uses a dynamic spatial conditional $\beta$-convergence model to analyze trends in nitrogen oxide emissions.
Another key focus is maternal and child health, particularly low birth weight and disparities in birth rates. These studies examine how socioeconomic, demographic, climatic, and healthcare-related factors interact with spatio-temporal risk patterns \cite{lome-hurtado_patterns_2021, ma_bayesian_2023, wang_spatio-temporal_2025}. Methodological innovation is also a motivation in this field. For example, combining survey data within a spatio-temporal framework enables more precise estimates and facilitates the identification of evolving hotspots in female genital mutilation prevalence \cite{nnanatu_evaluating_2021}. In all these cases, regression models are used exclusively with hierarchical structures.
A third application area is food security and nutrition. Spatio-temporal models are used here to identify dependencies and inform targeted policy interventions \cite{bofa_optimizing_2024, bofa_bayesian_2024}. While \cite{bofa_optimizing_2024} applies a hierarchical regression model, \cite{bofa_bayesian_2024} uses a flat model architecture.

\textit{Model strategies.}
We provide an overview in Table \ref{tab:models-publichealth}. In total, seven models employ a hierarchical architecture, and four use a flat architecture. All ten models incorporate an additive spatio-temporal structure (see Section  \ref{subsec:ST-CoV}). Two of these models additionally include lag structures (see Section  \ref{subsec:lag-structures}).
\begin{table}[h]
\footnotesize
\begin{tabularx}{\linewidth}{p{3cm}cX}
\hline
                        & Frequency  & References \\ \hline
  Hierarchical          &   6      &  \cite{bofa_optimizing_2024, lome-hurtado_patterns_2021, ma_bayesian_2023, nnanatu_evaluating_2021,  saez_spatial_2022, wang_spatio-temporal_2025}\\ 
  Flat                  &   4      & \cite{beloconi_spatio-temporal_2021, bofa_bayesian_2024, he_identifying_2021, martenies_spatiotemporal_2021}\\ \hline
 Additive Structures          &     10    &  \cite{beloconi_spatio-temporal_2021, bofa_optimizing_2024, bofa_bayesian_2024, he_identifying_2021, lome-hurtado_patterns_2021, ma_bayesian_2023, martenies_spatiotemporal_2021, nnanatu_evaluating_2021, saez_spatial_2022, wang_spatio-temporal_2025}\\ 
  Lag Structures     &   2      &  \cite{beloconi_spatio-temporal_2021, he_identifying_2021}\\ \hline
\end{tabularx}
\caption{Models classified according to the proposed scheme in the application domain public health}\label{tab:models-publichealth}
\end{table}

\textit{Limitations.}
Data quality and availability frequently constrain model accuracy. Several studies rely on limited historical records \cite{martenies_spatiotemporal_2021}, face general data availability issues \cite{bofa_bayesian_2024, he_identifying_2021}, or work with aggregated data at coarse spatial resolution \cite{lome-hurtado_patterns_2021}. Others depend on self-reported data \cite{nnanatu_evaluating_2021}, which may introduce bias or obscure fine-scale heterogeneity.
Model assumptions and simplifications also present challenges. Simplified representations, such as binary lockdown indicators \cite{beloconi_spatio-temporal_2021}, insufficiently differentiated variables \cite{he_identifying_2021}, or assumptions of no population mobility \cite{lome-hurtado_patterns_2021}, can overlook heterogeneity in exposures or interventions, limiting causal interpretability. The use of proxies (e.g., black carbon as a surrogate for traffic-related air pollution \cite{martenies_spatiotemporal_2021}) or assumptions about the independence or separability of spatial and temporal correlations \cite{saez_spatial_2022} may also compromise model realism.
Computational and methodological constraints further restrict study design and implementation. Some models require pre-specification of the number of clusters \cite{ma_bayesian_2023} or rely heavily on informative Bayesian priors \cite{bofa_optimizing_2024, bofa_bayesian_2024}, which may influence outcomes. Additionally, unaccounted variables such as omitted risk factors \cite{lome-hurtado_patterns_2021} or short-term events \cite{wang_spatio-temporal_2025} may distort the results.

%%%%%%%%%%%%%%%%%%%%%%%%%%%%%%%%%%%%
\subsection{Economics.}\label{subsub:Economics}
We classify six of the reviewed papers as belonging to the field of economics (see Table \ref{tab:appl.areas}).

\textit{Purpose of model fitting.}
A central motivation in economic applications is to understand productivity and growth patterns across regions. For instance, agricultural productivity is studied with particular attention to the scale, scope, and nature of spatial dependence \cite{baldoni_agricultural_2021}. At the same time, other research investigates how regional disparities influence national economic performance \cite{pietak_effect_2024}. Spatio-temporal models are also employed to examine interconnections between output growth and unemployment across provinces \cite{elhorst_spatial_2022}, or to forecast house price growth rates \cite{mattera_forecasting_2025}.
Beyond productivity, spatio-temporal frameworks are used to analyze tourism and urban systems. In tourism, such models assess destination competitiveness and resilience to economic shocks by capturing spatio-temporal flow patterns \cite{costantino_spatial_2023}. In urban economics, spatial structures are examined to uncover nonlinear effects on green economic efficiency and spatial spillovers \cite{yan_how_2025}.

\textit{Model strategies.}
Table \ref{tab:models-economics} provides an overview. All reviewed models in the economics domain follow a flat model architecture. Each model incorporates lag structures (see Section  \ref{subsec:lag-structures}), and five out of six also employ additive spatio-temporal structures (see Section  \ref{subsec:ST-CoV}), whereby all of the additive components are specified through fixed effects. The majority are spatial panel data models, with three studies specifically using dynamic spatial panel data models \cite{baldoni_agricultural_2021, costantino_spatial_2023, elhorst_spatial_2022}. One study uses a panel vector auto-regression model \cite{mattera_forecasting_2025}. 
\begin{table}[h]
\footnotesize
\begin{tabularx}{\linewidth}{p{3cm}cX}
\hline
                        & Frequency  & References \\ \hline
  Hierarchical          &    0    &  \\ 
  Flat                  &    6    & \cite{baldoni_agricultural_2021, costantino_spatial_2023, elhorst_spatial_2022, mattera_forecasting_2025, pietak_effect_2024, yan_how_2025}\\ \hline
  Additive Structures          &   5      & \cite{baldoni_agricultural_2021, costantino_spatial_2023, elhorst_spatial_2022, pietak_effect_2024, yan_how_2025} \\ 
  Lag Structures     &   6     &  \cite{baldoni_agricultural_2021, costantino_spatial_2023, elhorst_spatial_2022, mattera_forecasting_2025, pietak_effect_2024, yan_how_2025} \\ \hline
\end{tabularx}
\caption{Models classified according to the proposed scheme in the application domain economics}\label{tab:models-economics}
\end{table}

\textit{Limitations.}
Data limitations are a recurrent challenge. Issues such as missing covariates, coarse spatial aggregation, and the absence of alternative data sources restrict the exploration of spatial heterogeneity and hinder the inclusion of relevant explanatory variables \cite{pietak_effect_2024, yan_how_2025}.
Modeling assumptions and specifications introduce further constraints. Many studies rely on predefined spatial weight matrices \cite{baldoni_agricultural_2021, pietak_effect_2024}, geographical distance-based structures \cite{mattera_forecasting_2025}, or assume temporal stationarity, all of which may fail to capture dynamic spatial dependencies. Simplifications may reduce interpretability, such as focusing solely on morphological rather than functional or network-based spatial linkages \cite{yan_how_2025}, omitting boundary regions \cite{elhorst_spatial_2022}, or assuming global factors dominate over local cluster-specific dynamics \cite{mattera_forecasting_2025}.
Additional challenges include methodological difficulties in empirically distinguishing production fundamentals from productivity spillovers \cite{baldoni_agricultural_2021}, as well as computational limitations in handling complex model structures \cite{elhorst_spatial_2022}.

%%%%%%%%%%%%%%%%%%%%%%%%%%%%%%%%%%%%
\subsection{Criminology.}\label{subsub:Crime}
We also assign six publications to the criminology category (see Table \ref{tab:appl.areas}).

\textit{Purpose of model fitting.}
One motivation for applying spatio-temporal models in crime research is to address data limitations and uncertainty, particularly regarding the imprecise timing of crime events \cite{briz-redon_bayesian_2024}. 
Another central aim is forecasting crime patterns to support prevention strategies and resource allocation. Models are used to predict burglary frequencies and inform police deployment \cite{mahfoud_forecasting_2021}, while modern modeling approaches are compared to classical ones for their interpretability and practical relevance \cite{clark_class_2021}.
A further line of research seeks to understand violence and its contextual drivers. For example, studies examine neighborhood-level influences on intimate partner violence \cite{gracia_chronic_2021}, or jointly model different forms of violence against women to identify shared spatio-temporal structures \cite{vicente_multivariate_2023}.
Finally, some research explores spillover effects in crime data, not only in terms of offender behavior but also reporting dynamics \cite{vomfell_no_2023}.

\textit{Model strategies.}
An overview is provided in Table \ref{tab:models-crime}. All models in the criminology domain follow a hierarchical architecture. The majority are regression models with additive spatio-temporal structures (see Section  \ref{subsec:ST-CoV}). One model \cite{gracia_chronic_2021} additionally incorporates lag structures (see Section  \ref{subsec:lag-structures}) to capture spatio-temporal dependencies. We also find the spatially correlated self-exciting model in \cite{clark_class_2021}, and the Hawks process in \cite{vomfell_no_2023}.
\begin{table}[h]
\footnotesize
\begin{tabularx}{\linewidth}{p{3cm}cX}
\hline
                        & Frequency  & References \\ \hline
  Hierarchical          &   6     &  \cite{briz-redon_bayesian_2024, clark_class_2021, gracia_chronic_2021, mahfoud_forecasting_2021, vicente_multivariate_2023, vomfell_no_2023}\\ 
  Flat                  &   0     & \\ \hline
  Additive Structures       &    4     & \cite{briz-redon_bayesian_2024, gracia_chronic_2021, mahfoud_forecasting_2021, vicente_multivariate_2023} \\ 
  Lag Structures     &    1     & \cite{gracia_chronic_2021}  \\ 
  Intensity Function &   1    &  \cite{vomfell_no_2023} \\ 
  Others: &  & \\
  \hspace{0.05cm} self-exciting model & 1 & \cite{clark_class_2021} \\\hline
\end{tabularx}
\caption{Models classified according to the proposed scheme in the application domain criminology}\label{tab:models-crime}
\end{table}

\textit{Limitations.}
Data limitations are prominent, as many studies rely on official records that may not reflect the full scope of criminal activity. For example, reported cases of intimate partner violence capture only severe incidents, underestimating true prevalence \cite{gracia_chronic_2021}. Similarly, assumptions about spatial proximity or aoristic temporal distributions may not fully capture underlying social structures or event timing \cite{briz-redon_bayesian_2024, vomfell_no_2023}.
Simplifying assumptions, such as using logistic regression for temporal uncertainty or assuming purely local spillovers, may limit model realism \cite{briz-redon_bayesian_2024, clark_class_2021}. In multivariate frameworks, the number of parameters grows rapidly with the number of crime types, increasing complexity and potential overfitting \cite{vicente_multivariate_2023}.
Computational constraints are another major issue. High computational costs, especially with complex Bayesian or INLA-based approaches, restrict exploratory analyses and limit the inclusion of fine-grained covariates or large datasets \cite{clark_class_2021, mahfoud_forecasting_2021}.
Generalization is also limited, as many models are developed for specific urban areas and may not extend to rural regions or other cultural contexts. Additionally, some neighborhood-level mechanisms, such as social norms or collective efficacy, are often unobserved, restricting the ability to test specific theoretical relationships \cite{gracia_chronic_2021}.
\newpage

\textbf{Funding Information}

This work was supported by a fellowship of the German Academic Exchange Service (DAAD). This work was partially supported by JSPS KAKENHI Grants JP21H04907 and JP24H00732, by JST CREST Grant JPMJCR20D3 and JPMJCR2562 including AIP challenge program, by JST AIP Acceleration Grant JPMJCR24U3, and by JST K Program Grant JPMJKP24C2 Japan.

\textbf{Acknowledgments}

The authors would like to thank Christoph Mandl, Florian Braun, and Filip Kučera for their constructive feedback and helpful discussions.

\textbf{Data Availability}

Data sharing is not applicable to this article as no new data were created or analyzed in this study.

\textbf{Notes}

We would like to point out that the authors used ChatGPT during the writing process of this paper, primarily for minor rephrasing and grammar corrections.

\textbf{Conflict of Interest}

The authors declare that they have no conflicts of interest.

\printbibliography

%\textbf{Further Reading}

\end{document}